\documentclass[12pt,american,1pt, review]{elsarticle}
\usepackage[LGR,T1]{fontenc}
\usepackage[latin9]{inputenc}
\usepackage{array}
\usepackage{wrapfig}
\usepackage{textcomp}
\usepackage{amsmath}
\usepackage{amssymb}
\usepackage{graphicx}
\usepackage{setspace}
\usepackage{subscript}
\makeatletter

\DeclareRobustCommand{\greektext}{%
  \fontencoding{LGR}\selectfont\def\encodingdefault{LGR}}
\DeclareRobustCommand{\textgreek}[1]{\leavevmode{\greektext #1}}
\ProvideTextCommand{\~}{LGR}[1]{\char126#1}

\providecommand{\tabularnewline}{\\}

\usepackage{amsmath,epsfig,amssymb,graphicx}
\usepackage{bbm}
\usepackage{setspace,exscale,latexsym,srcltx}
\usepackage{longtable,pdflscape}
\usepackage[scanall]{psfrag}
\usepackage{rotating}
\usepackage[nottoc]{tocbibind}
\usepackage{dsfont}

\usepackage{tikz}
\usepackage{subcaption}

\usepackage{natbib}
\usepackage{breakurl} 

\usepackage{url}

\usepackage{rotating}
\usepackage{tocloft}
\usepackage{tikz}
\usepackage{color}
\usepackage{tocbibind}
\usepackage{changepage}

\usepackage[colorlinks=true,linkcolor=blue,citecolor=blue]{hyperref}%
\usepackage[labelfont=bf]{caption}

\usepackage[english]{babel}
\usepackage{wrapfig}

\usepackage{booktabs}
 
\usepackage[margin=2.5cm]{geometry}%

\usepackage{har2nat}

\usepackage{fancyhdr}	
\allowdisplaybreaks

\makeatother

\usepackage{babel}
\begin{document}

\begin{frontmatter}{}

\title{{\Large{}Optimal Order Execution in Intraday Markets: Minimizing
Costs in Trade Trajectories}}

\author{{\normalsize{}Christopher Kath}\textsuperscript{a,{*}}}

\author{{\small{}Florian Ziel}\textsuperscript{{\small{}b}}}

\ead{christopher.kath@stud.uni-due.de}
\begin{abstract}
Optimal execution, i.e., the determination of the most cost-effective
way to trade volumes in continuous trading sessions, has been a topic
of interest in the equity trading world for years. Electricity intraday
trading slowly follows this trend but is far from being well-researched.
The underlying problem is a very complex one. Energy traders, producers,
and electricity wholesale companies receive various position updates
from customer businesses, renewable energy production, or plant outages
and need to trade these positions in intraday markets. They have a
variety of options when it comes to position sizing or timing. Is
it better to trade all amounts at once? Should they split orders into
smaller pieces? Taking the German continuous hourly intraday market
as an example, this paper derives an appropriate model for electricity
trading. We present our results from an out-of-sample study and differentiate
between simple benchmark models and our more refined optimization
approach that takes into account order book depth, time to delivery,
and different trading regimes like XBID (Cross-Border Intraday Project)
trading. Our paper is highly relevant as it contributes further insight
into the academic discussion of algorithmic execution in continuous
intraday markets and serves as an orientation for practitioners. Our
initial results suggest that optimal execution strategies have a considerable
monetary impact.
\end{abstract}

\address{\textsuperscript{a}University Duisburg-Essen, Chair for Environmental
Economics}

\address{\textsuperscript{b}University Duisburg-Essen, Chair for Environmental
Economics}

\ead{florian.ziel@uni-due.de}
\begin{keyword}
Intraday electricity market, Market Microstructure, Optimal Execution,
Algorithmic Trading

\cortext[cor1]{Corresponding author}
\end{keyword}

\end{frontmatter}{}

\section{Introduction}

Intraday markets are of tremendous importance in German electricity
trading. Traded volumes rise from year to year and set a new record
in 2019 (\citet{EPEXvolcite}). The growing share of renewable energy
requires market participants to balance their positions in very short-term
intraday markets and causes more traded volume close before delivery
(\citet{koch2019short}). Intraday trading is usually used for rebalancing
after power plant outages, updated consumption forecasts, or speculative
trading approaches, as demonstrated by \citet{maciejowska2019day}.
This paper refers to hourly continuous intraday trading, leaving quarter-hourly
trading and intraday auctions aside. The German market allowed for
trading up to five minutes before delivery in mid-2020 coupled with
other European intraday markets, as \citet{kath2019modeling} analyzed
and worked on a pay-as-bid basis, meaning that orders were continuously
matched and executed at their respective prices. Figure 1 displays
the different time lines of German hourly intraday trading, including
the coupling among markets under the Cross-border intraday project
(XBID). More information on the German market is also provided by
\citet{viehmann2017state}.\\
\hspace*{0.5cm}While the German intraday market itself has develoed
into a central market-place, its many simultaneously traded hours
have also brought another topic to the agenda: automated and algorithmic
trading. The term describes the algorithm-aided execution of trades
or liquidity provision strategies, as mentioned by \citet{hendershott2013algorithmic}.
Herein, we will drill down our focus to optimal execution and leave
liquidity provision strategies aside. The continuous matching process
is crucial in understanding the main issue of optimal execution. Energy
companies face a practical problem: Their trading books are flooded
with volumes, which may stem from outages, renewable generation, or
trading decisions. But how can they execute the volumes in an optimal
way? A market participant could trade the entire volume at once, wait
for a better point in time with more favorable prices, or even slice
the volume into many small portions to be traded. The slicing aspect,
or the determination of an optimal trading trajectory, is what optimal
execution aims to identify.\\
\begin{figure}
\centering{}\includegraphics[scale=0.65]{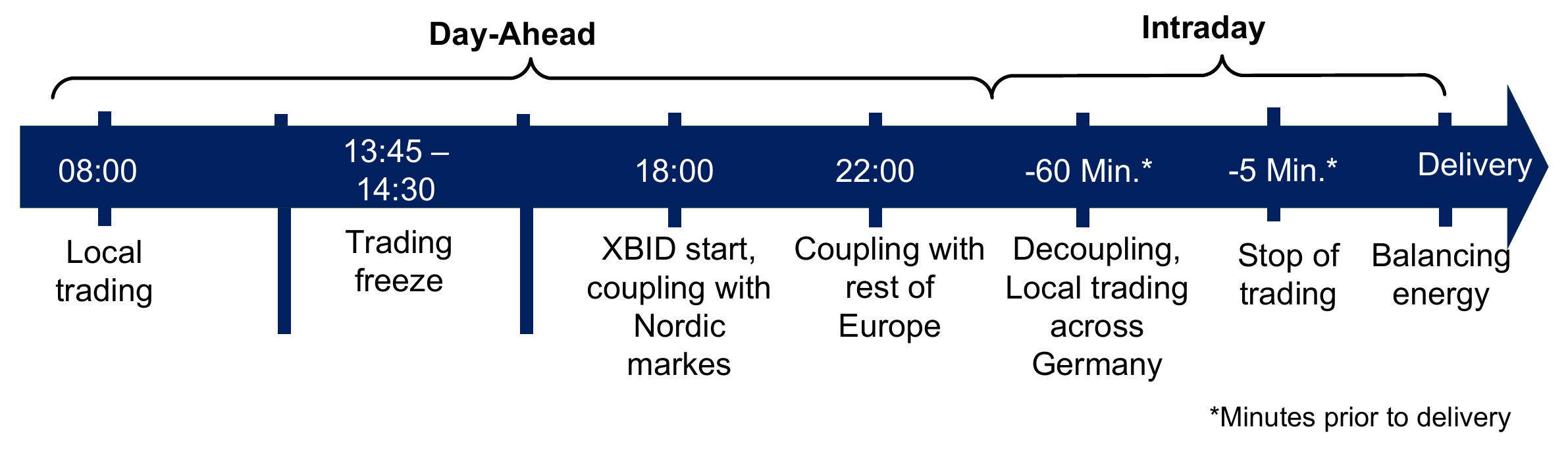}\caption{Chronological order of the German hourly intraday market.}
\end{figure}
There is a rich body of literature on analytical aspects that discuss
market determinants, such as \citet{kiesel2017econometric,hagemann2015price,narajewski2018econometric,pape2016fundamentals},
and forecasting approaches, like \citet{uniejewski2019understanding,janke2019forecasting,kath2018value},
and \citep{kath2018value}. From a practical perspective, these papers
only discuss reasons for the price movements and predictability of
intraday trading but leave the question of 'how to trade' open. The
problem of executing orders in the most optimal way was referred to
by \citet{garnier2015balancing} and \citet{aid2016optimal} with
a strict focus on renewable energy source (RES) generation. However,
what the current academic discussion misses is a more generic discussion
of optimal execution strategies in the German electricity intraday
market. While there are many stochastic models under the umbrella
of equity trading (e.g., \citet{almgren2001optimal,almgren2003optimal,bertsimas1998optimal}),
there were only three electricity trading approaches present when
this paper was written. The first one, provided by \citet{von2017optimal},
discusses optimal market-making strategies in the context of the German
intraday market. Meanwhile, \citet{glas2019intraday} and \citet{glas2020intraday}
addressed a more general numerical execution approach for both renewable
and conventional generation output. Finally, \citet{Coulon} presented
a preliminary model on executing volumes of renewable generation in
an optimal manner.\\
\hspace*{0.5cm}We aim to position the topic of optimal execution
more prominently in the current discussion of German intraday markets
and analyze various approaches in the remainder of this paper. Our
work is structured as follows: Section \color{blue} \ref{sec:2} \color{black}
takes a deeper look at order books and the underlying data. Most of
the time, actual trades are analyzed in the context of intraday trading
which is why we need to address what distinguishes order book data
from trades. In the course of the data discussion, we will derive
the problem of optimal execution in a more formal way. Section \color{blue} \ref{sec:3} \color{black}
introduces execution approaches and discusses, which weaknesses they
exhibit in the context of electricity markets. Leaving the theoretical
part aside, we compute trading trajectories under different scenarios
and evaluate the results in an empirical out-of-sample study in Section
\color{blue} \ref{sec:4} \color{black} and summarize our findings
and the contributions of this paper in Section \color{blue} \ref{sec:5}. \color{black}

\section{Order Book Data and the Problem of Optimal Execution\label{sec:2}}

\subsection{Why Order Book Data are Different}

\begin{figure}
\centering{}\includegraphics[scale=0.5]{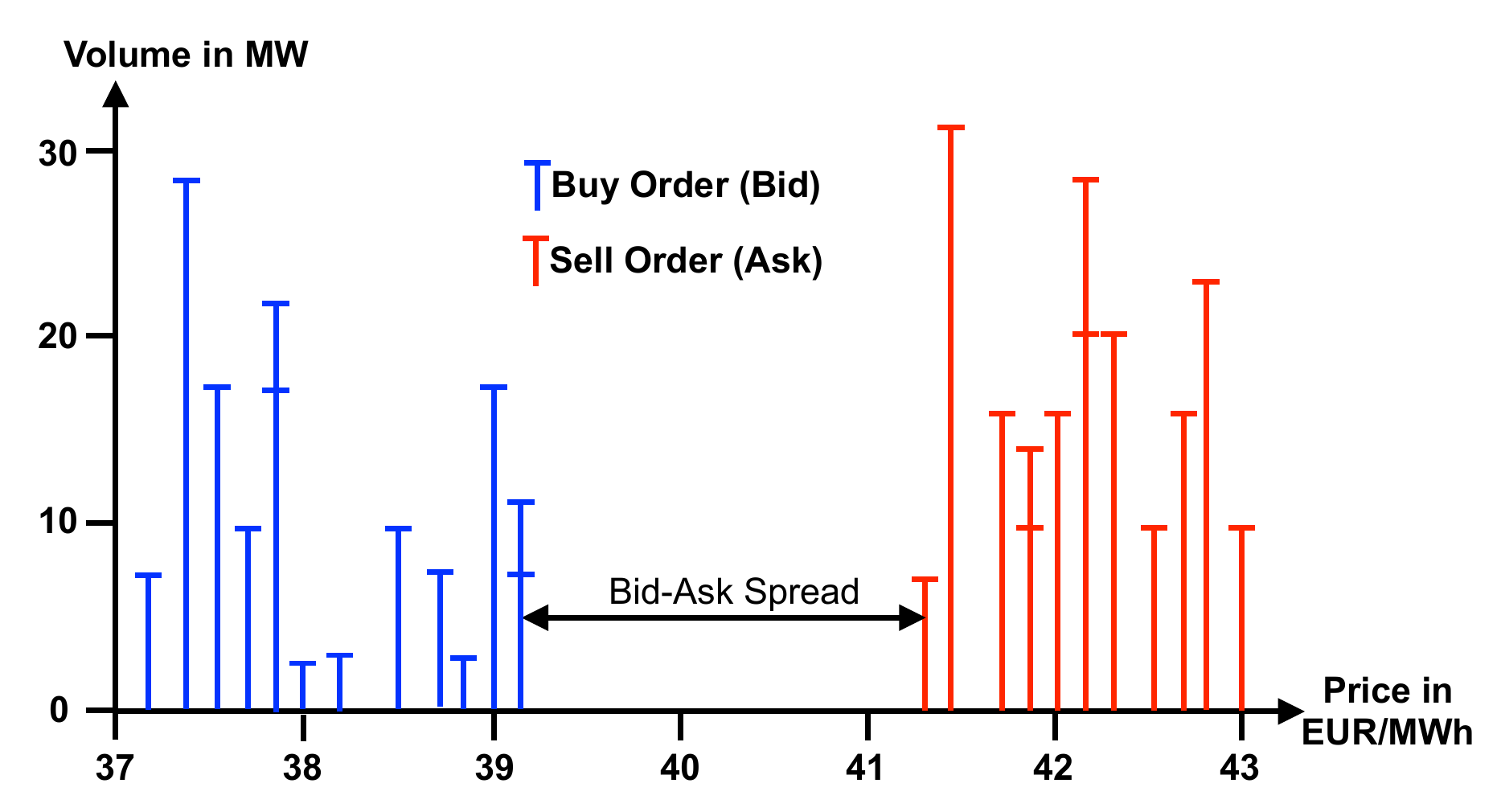}\caption{Schematic depiction of a typical intraday continuous order book for
a specific delivery hour. The best bid is the order that is willing
to buy at the highest price, while the best ask is that ready to sell
at the lowest price. Please note that there can be multiple orders
at the same price level, as shown at the best bid order.}
\end{figure}
If one takes a historic view of intraday papers, the evolution of
complexity and availability in the research data becomes evident.
Early articles like \citet{hagemann2015price} solely used volume-weighted
average prices of the entire trading period. This modus operandi ignores
the characteristics of continuous trading and takes a more time series-oriented
approach. Later papers dealt with individual intraday trade data (e.g.,
\citet{janke2019forecasting}). They focused on individual transactions
and derived partial averages. However, this is only one side of the
coin since these papers focused on trades, i.e., the buying and selling
of orders that could be matched. Better availability of data has allowed
researchers to analyze order book data. These data-sets comprise not
only trades but also unmatched orders. Orders grant new insight into
continuous trading and the market microstructure behind it.\\
\hspace*{0.5cm}In a limit order book (LOB), orders are continuously
sorted based on the price per buying and selling direction for each
delivery hour. Thus, a typical LOB for one delivery hour can be imagined
like a t-account with buy and sell sides. Figure 2 represents an order
book in a graphical manner. All buy orders, or the \textquotedbl bid\textquotedbl{}
side, are sorted in descending order, while all sell orders, or the
\textquotedbl ask\textquotedbl{} side, are ordered in ascending order,
such that the best sell order in the market is the one with the lowest
price.\\
\hspace*{0.5cm}
\begin{figure}
\centering{}\includegraphics[scale=0.5]{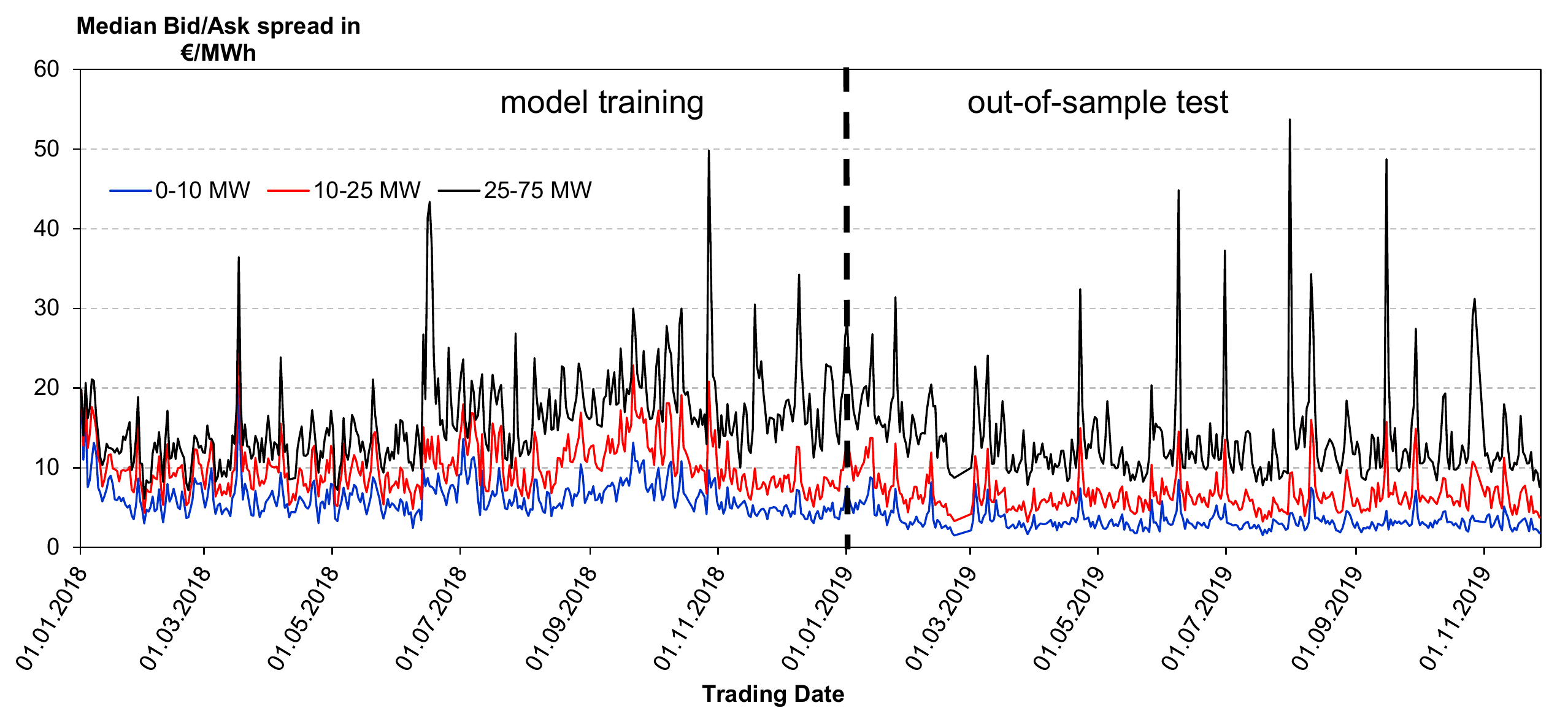}\caption{Median bid-ask spread from January 2018 to November 2019. The data
were aggregated in minute buckets and reflect the development of spreads
under the trading volume, i.e., the typical cost of trading averaged
for one minute of order book activity for the first 0 to 10 MW, 10
to 25 MW, and the next 25 to 75 MW. The data are separated into a
model training phase and an out-of-sample section to realistically
benchmark the execution models.}
\end{figure}
The next crucial aspect of understanding LOB data is timing. Figure
2 displays a snapshot of one specific second. However, we need to
define a describing element that discretizes countless snapshots and
aggregates the information. The bid-ask spread (BAS) serves this aspect
well as it compresses buy and sell orders into one numerical figure.
However, LOB data are not one-dimensional. Figure 2 reveals the connection
of volumes available at different price levels. It makes no sense
to focus on the best bid and ask for the BAS computation if the volume
sums up to 0.1 MWh. Therefore, we aggregate the BAS into different
volume buckets, i.e., BAS per first 1 MW traded, BAS for the volume
between 1 and 5 MW, and so on (see Section 2.2 for the mathematical
formulation). Last but not least, we need a second level of discretization
to cover the aspect of time. We can aggregate our data into minute
buckets such that we can identify the best 1 MW BAS over an interval
of one minute of trading, for instance.\\
\hspace*{0.5cm}We used LOB data from 01.01.2019 to 27.11.2019 provided
by EPEX Spot SE. The data can be purchased from EPEX Spot SE. For
more information on EPEX order book data and the required data preparation,
refer to \citet{martin2018german}. Unfortunately, the data are not
completely available for the year 2019, which is why Figure 3 ends
in late November 2019. We also split the data into training (year
2018) and out-of-sample testing (year 2019). Some execution algorithms
require historic data for model training. We used the entire year
of 2018 for model training. All data of 2019 were used to benchmark
the models in an out-of-sample manner.\\
\hspace*{0.5cm}The series of median BAS data, as shown in Figure
3, does not suggest any yearly seasonality. There are some spikes,
but, all in all, the spreads seem to be stable across the year. Figure
4 zooms in and shows the median BAS per weekday and delivery hour
as well as the corresponding traded volume. A similar pattern evolves
for both the weekly and hourly bid-ask series. They appear to be stable
no matter the day or delivery hour. The market participants seem to
have similar order quoting behavior. This is an important first observation
as it leads to a conclusion on execution algorithms. If quoting is
stable, considering seasonality is not a major concern. The picture
changes with the trading volume. Saturdays and Sundays feature lower
trading volumes. Figure 4 delivers evidence for hourly patterns: The
delivery hours 1\textemdash 7 show signs of less trading volume.\\
\hspace*{0.5cm}Last but not least, we want to analyze the continuous
trading character of German intraday markets. Figure 5 shows how the
median BAS behaves with respect to time to delivery. Its level is
constant over time but changes 60 and 30 minutes before delivery.
This effect is connected to the local coupling of markets. One hour
before delivery, no European orders are coupled under XBID with the
German market anymore, and 30 minutes before delivery, local within-grid
area trading starts (see \citet{kath2019modeling}). Consequently,
volumes sharply drop at 60 and 30 minutes to recover later. Figure
5 is also a good summary of the problem of optimal execution itself.
How should volumes be distributed over the available trading time
up to delivery? Moreover, how should volumes be sliced to ensure that
a minimum spread is paid? 

\subsection{Discretization Methodology }

The amount of data with LOB requires different handling than that
with trade data. Our dataset is around 60 GB, it does not allow for
simple calculation on an order basis anymore but demands an aggregation
approach. Orders need to be aggregated by \textit{time} and \textit{trading
volume}. We will start with \textit{time}. To understand the concept
of time aggregation, a few notations need to be introduced. Imagine
one line in our data frame as 
\begin{equation}
Z:\left\{ y_{t,d},v_{t,d},s_{t,d},e_{t,d},d_{t,d}|t=1,...,T\wedge d=[\mathrm{buy,sell}]\right\} .
\end{equation}
Let this be the set of all buy and sell orders at delivery time $t$
comprising its price $y_{t,d}$, the volume of an order denoted as
$v_{t,d}$, its start $s_{t,d}$, and expiry time stamp $e_{t,d}$
in seconds from the start of the observation. We divide the difference
between the reception of any position to be traded at $i$ and the
delivery at time $t$ in equidistant steps $k=1,...,N$, where $N=\left\lfloor \frac{t}{(24\text{\texttimes}60\text{\texttimes}60)}-\frac{i}{(24\text{\texttimes}60\text{\texttimes}60)}\right\rfloor $.
Thus, we have $k$ buckets of one minute in length. The parameter
$k$ is a dynamic one depending on the arrival of the position at
$i$. That means that if the time of arrival changes, the computation
of the optimal execution likewise differs, as the length of $k$ is
different. Based on $k$, it is possible to define time discretization
buckets $O_{t,k,d}$
\begin{equation}
O_{t,k,d}=\left\{ y_{t,d},v_{t,d},s_{t,d},e_{t,d},d_{t,d}\in Z|(s<60\times k\times1\wedge e\geqslant60(k-1))\right\} ,
\end{equation}
which comprise all active orders\footnote{Note that we need to remove cancellations from the set of all active
orders in order to avoid counting any orders twice.} in the $k$-th minute interval bucket for each delivery time stamp
$t$ and direction $d$. Note that this automatically filters out
unnecessary information present before the position arrival at $i$,
which reduces the memory space used. The choice of minute buckets
seems suitable for the EPEX intraday market as with lower granularity,
one faces many instances with missing data as there are simply no
new activities, e.g., one will face many buckets with no trades every
second. However, decreasing the resolution to 15-minute buckets lessens
the trading abilities of the algorithms and might turn out to be a
poor choice, as the volume is split on fewer buckets with less trades
but more volume per trade and, as a consequence, higher spreads. Therefore,
one-minute buckets appear to be a good compromise. Other approaches
like \citet{glas2020intraday} use five minute intervals, but we believe
that a finer grid causes the results to be more realistic.\\
\hspace*{0.5cm}As a next step, we need to filter by cumulative sum
$CS_{t,d}$ of volume $v_{t}$ ordered by price $y_{t,d}$ in an ascending
(in the case of buy orders) or descending manner (in the case of sell
orders). First, we define auxiliary volume buckets $b_{r}$ with $r=1,...,23$
for our aggregation logic as $b_{1}=[0,1],$$b_{2}=[1,5],$$b_{3,..,7}=[5,10],...,[25,30]$,
$b_{8,..,14}=[30,40],...,[90,100]$, $b_{15,..,18}=[100,125],...,[175,200]$,
$b_{19,20}=[200,250],[250,300]$, $b_{21,22}=[300,400],[400,500]$
and $b_{23}=[500,\infty]$. This leads to a refined version of Eq.
(2) that allocates not only based on the $k$-th time bucket but also
on the $r$-th volume bucket in

\begin{equation}
O_{t,k,r,d}=\left\{ y_{t,d},v_{t,d},s_{t,d},e_{t,d},d_{t,d}\in Z|(s<60\times k\times1\wedge e\geqslant60(k-1))\wedge CS\in b_{r}\right\} .
\end{equation}
The buckets $b_{r}$ serve as separators for the cumulative volume.
Obviously, trading costs vary over time. However, they also change
with the designated trading volume, as depicted in Figure 5. The LOB
is ordered by price in the case of both buying and selling. Unless
all orders are entered with identical prices, a price ladder will
emerge. The best buy and sell prices might feature only a small share
of the total volume. As a consequence, trading usually gets more expensive
with large quantities. We can take this aspect into consideration
by aggregating buckets by volume as well. \\
\hspace*{0.5cm}
\begin{figure}
\centering{}\includegraphics[scale=0.45]{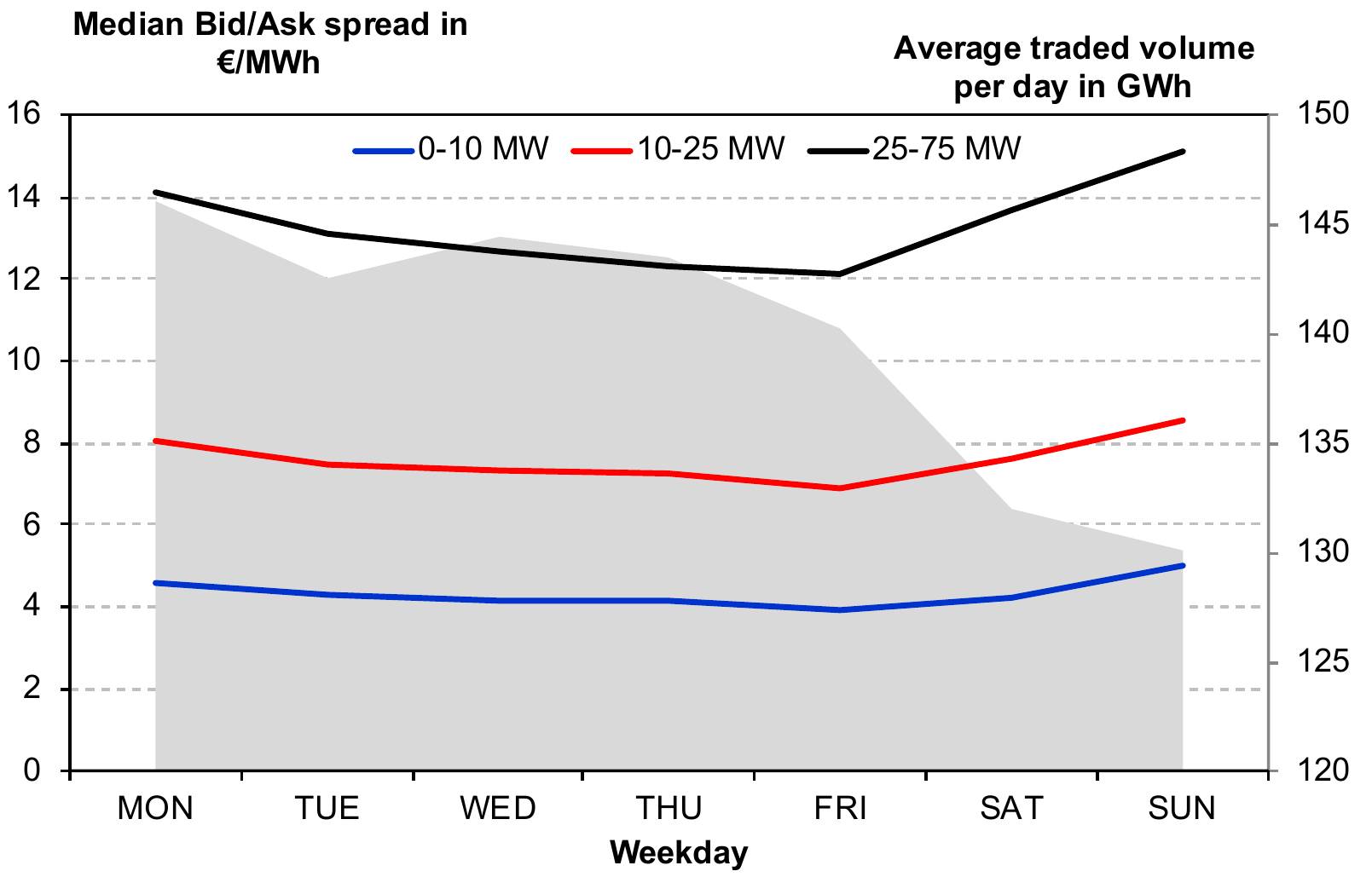}\includegraphics[scale=0.45]{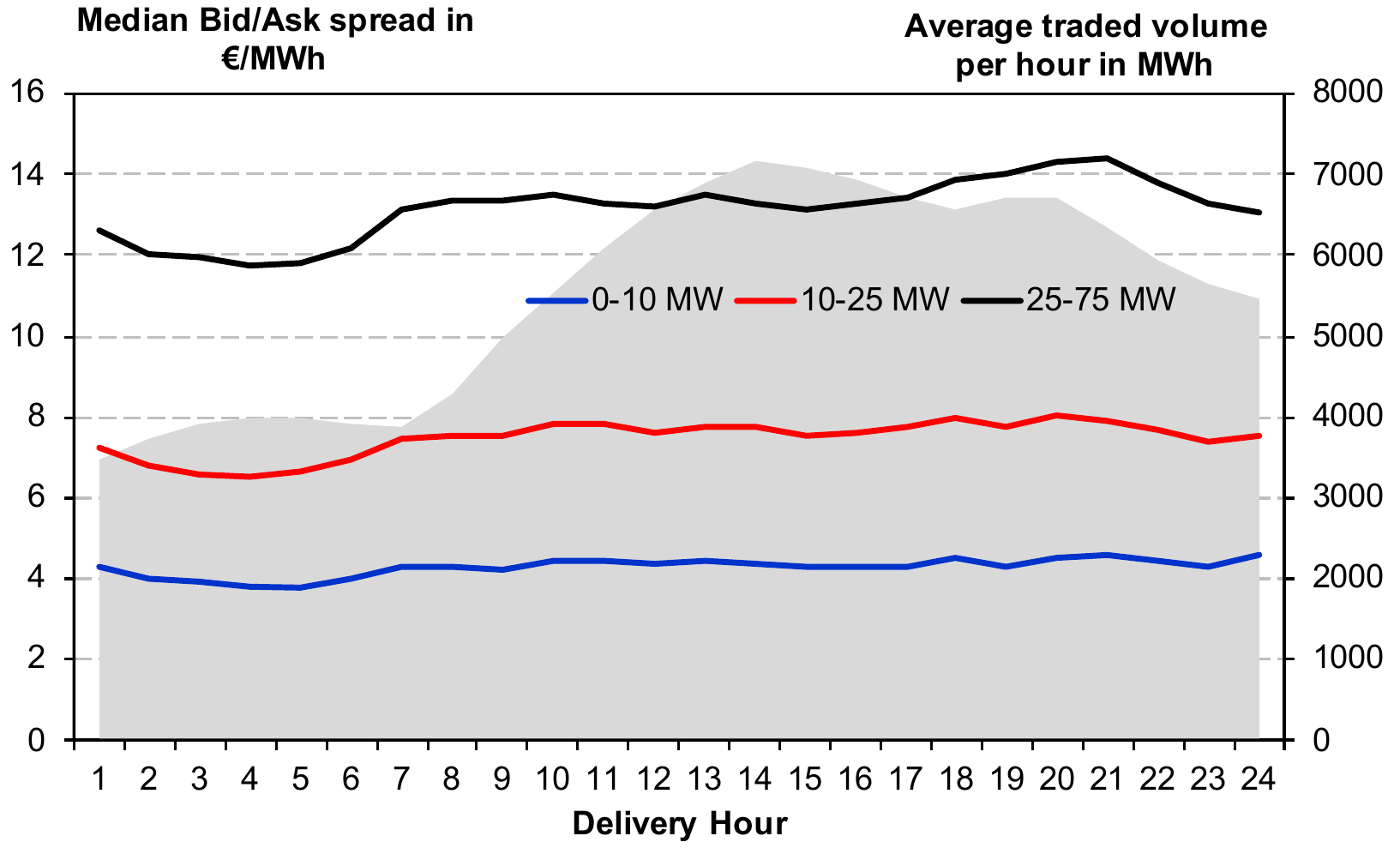}\caption{Evolvement of weekly and hourly patterns in median BAS and volumes
of aggregate minute clips ranging from January 2018 to November 2019.}
\end{figure}
One could argue that the aspect of time-weighted orders is missing
in this approach. It makes a difference if an order exists for a few
seconds or for the entire bucket range of 60 seconds. We have neglected
this aspect, as the additional volume weighting ensures that small
orders that just exist for a few seconds do not count too much. A
limited simulation showed that there were no substantial differences
if we added another time weighting component, which is why we have
left it out for the sake of a faster computational time. 

\subsection{Underlying Market Assumptions and Problem Formulation}

The market microstructure of intraday markets requires more consideration
as well. Therefore, we assume the following for our optimal execution
analysis:
\begin{itemize}
\item The algorithms can only accept existing orders in the LOB and actively
pay the BAS. To understand this concept, we need to discuss the two
execution alternatives. A market player can initiate an order by quoting
a price and volume pair. Say, for instance, the best buy order is
at 50 €/MWh, and the best sell order is at 51 €/MWh. A trader can
initiate another buy at 50.10 €/MWh and wait for execution, Or they
can actively accept an existing order by entering a buy at 51 €/MWh,
which results in immediate execution as there is a corresponding sell
order available. Technically speaking, the latter can be seen as an
active click in the LOB. We assume it as the only way to trade, since
waiting for other traders to accept the order is very complex to depict.
It requires modeling other traders who are willing to accept the order.
\item We ignore further trading costs (such as fees), as they are usually
constant in time and volume and should not influence our optimal execution
path. In addition, they are counter-party specific and vary based
on bilateral agreements between traders, the exchange, and clearing
banks.
\item The optimal execution is probed for a fixed trading volume. This is,
for instance, the case for plant outages that need to be covered in
the intraday market. Generation updates for renewable energy sources
imply uncertainty in forecasts and changing position sizes, which
add another dimension of complexity and are ignored in this paper.
\item Following \citet{narajewski2018econometric}, the intraday market
is considered as efficient in this paper. Thus, we do not have any
view or opinion on price developments and only focus on execution.
\item All algorithms stop their activities 30 minutes before delivery. \citet{kath2019modeling}
showed that this is the time when intra-control area trading starts.
Therefore, it makes sense to see this threshold as an operational
barrier, as no Germany-wide trading is possible anymore. Apart from
that, the EPEX indices ID3 and ID1 stop 30 minutes before delivery
(for an ID3 example, see \citet{uniejewski2019understanding}), which
makes our approach time-congruent with the EPEX data. This implies
that, for instance, $k=180$ means 180 minutes of trading starting
from 210 minutes\footnote{It is important to understand the chronological differences. The trading
time stops 30 minutes before delivery. Thus, we can define trading
time=lead-time minus 30 minutes. } before delivery up to 30 minutes before delivery. 
\item We round all volumes stemming from the trading trajectories to one
digit whenever possible since this is the minimum tick size at the
exchange.
\end{itemize}
\begin{figure}
\centering{}\includegraphics[scale=0.5]{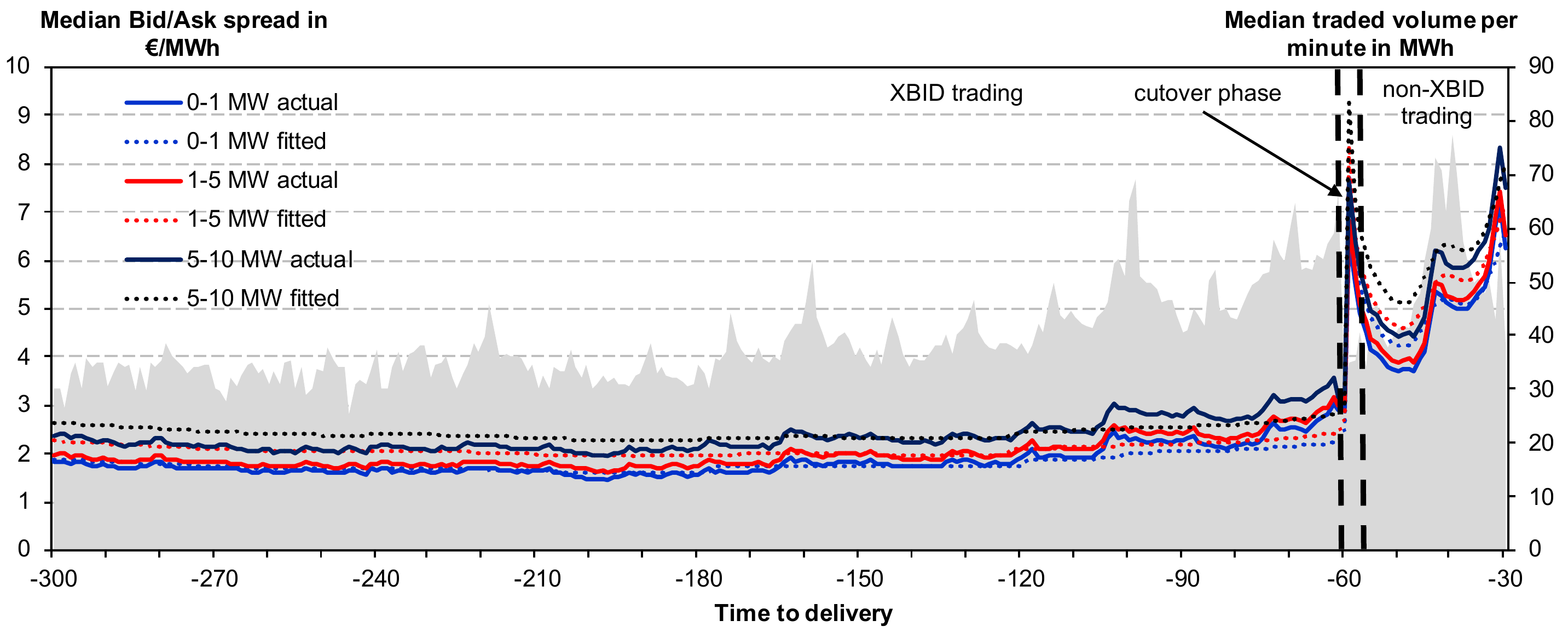}\caption{Median BAS and median volume of minute clips, i.e., all orders aggregated
every minute of trading with respect to time to delivery. The x-axis
shows the time in minutes until physical delivery. The dashed lines
reflect the in-sample fit of the spline-based market impact model
described in section 3.1 and emphasize how well our optimization model
can anticipate costs measured as median spreads.}
\end{figure}

\section{Optimal Execution Strategies\label{sec:3}}

\subsection{Proposed Execution Model Approach}

\subsubsection{Determinants of Optimal Execution}

The first model on optimal execution was provided by \citet{bertsimas1998optimal}.
Their contribution was a consideration of market impact, meaning the
temporary and constant change of prices due to own trading activity.
\citet{almgren2001optimal} expanded the idea by taking into account
the traders' risk appetite. They proposed a mean-variance-based approach
that locates an optimum between minimizing the trading costs and minimizing
the variance of costs. Later on, \citet{almgren2003optimal} expanded
the existing model with a non-linear market impact function. All models
share that they were applied on equity markets and have not been fully
tested in electricity intraday markets for optimal execution yet.
Herein, we borrow basic principles from the above mentioned papers
and derive a suitable model for intraday power markets.\\
\hspace*{0.5cm}First, we need to introduce some further notations
that are important for our model (see Table 1 for a detailed description
of the model parameters). We have
\begin{itemize}
\item a price $y_{k}$ (meaning a more general notation of prices here than
the one of Section 2.2, where we referred to the prices of individual
orders),
\item an overall amount of $X$ we need to sell or buy in the continuous
intraday market before its physical delivery at time $t$,
\item the volumes traded in each time step $k$ denoted as $n_{k}$ and
s.t. $X=\sum_{k=1}^{N}n_{k}$ and $x_{k}=X-\sum_{j=1}^{k}n_{j},$
\item the trade inventory position $x_{k}$ describing how much volume is
still to be traded at each time bucket $k$ (it follows that $x_{o}=X$
and $x_{N}=0$), and
\item the volatility of $y_{k}$ given by $\sigma_{k}$.
\end{itemize}
In the case of buying positions, $n_{k}$ shall be positive and in
the case of selling positions $n_{k}$ shall be negative. This assumption
allows us to have a position-neutral notation, as buying and selling
changes with the positive or negative position input of $n_{k}$.
\\
\hspace*{0.5cm}Another very important concept is the idea of market
impact. A common approach (e.g., in \citet{almgren2001optimal}) is
to divide it into two parts such that the market impact $I_{k}$ is
denoted by
\begin{equation}
I_{k}(n_{k},k,t)=I_{k}^{temp}(n_{k},k,t)+I_{k}^{perm}(n_{k-1},k-1,t),
\end{equation}
where $I_{k}^{temp}$ is the temporary component that is due to the
order book depth and less favorable prices that market participants
receive for large volumes; $I_{k}^{perm}$describes the permanent
change in market prices after trading has happened. It is important
to note that the permanent market impact is only observable in the
following epoch after the liquidity providers have re-entered their
orders, which is 
\begin{table}
\centering{}%
\begin{tabular}{c>{\centering}p{3cm}c>{\centering}p{5cm}}
{\footnotesize{}Determinant} & {\footnotesize{}Description} & {\footnotesize{}Value} & {\footnotesize{}Calculation/Derivation}\tabularnewline
\hline 
{\footnotesize{}$S_{0}$} & {\footnotesize{}Initial price of electricity {[}€/MWh{]}} & {\footnotesize{}47.22} & {\footnotesize{}Average price of all 2018 trades}\tabularnewline
{\footnotesize{}$X$} & {\footnotesize{}Total position to trade {[}MWh{]}} & {\footnotesize{}changing} & {\footnotesize{}To be adjusted per scenario}\tabularnewline
{\footnotesize{}$\sigma$} & {\footnotesize{}Daily volatility}{\footnotesize\par}

{\footnotesize{}{[}€/MWh{]}} & {\footnotesize{}20.57} & {\footnotesize{}Yearly volatility}\tabularnewline
{\footnotesize{}$\mu$} & {\footnotesize{}Annual growth}{\footnotesize\par}

{\footnotesize{}{[}€/MWh{]}} & {\footnotesize{}0} & {\footnotesize{}No price growth/drift is assumed}\tabularnewline
{\footnotesize{}$\lambda$} & {\footnotesize{}Risk aversion}{\footnotesize\par}

{\footnotesize{}{[}no specific unit{]}} & {\footnotesize{}2x10\textasciicircum (\textemdash 5)} & {\footnotesize{}Slightly less risk averse than \citet{almgren2001optimal}}\tabularnewline
\hline 
\end{tabular}\caption{Optimization model parameter used in the empirical study and the means
of derivation. The authors tried to estimate most of the parameters
based on empirical data from the training period in year 2018. However,
some parameters had to be guessed based on experience or applications
in the financial world.}
\end{table}
why its determinants $n_{k-1}$ and $k-1$ are lagged in Eq. (4).
The underlying price model, based on \citet{almgren2001optimal},
is assumed to follow
\begin{equation}
y_{k}=y_{k-1}+\sigma_{k}\xi_{k}-I_{k-1}^{perm}(n_{k-1},k-1,t),
\end{equation}
where $\xi_{j}$ is an independent random variable with $E[\xi_{k}]=0$
and i.i.d. as well as $Var[\xi{}_{k}]=1$. We acknowledge that these
assumptions are not totally perfect, as \citet{narajewski2020ensemble}
showed, but they seem to be reasonable for our application. \\
\hspace*{0.5cm}The permanent market impact refers to the most recent
time lag as the latest order is assumed to influence the current prices
in the form of its permanent market impact function. Note that we
have left out any drift term in Eq. (5). However, since electricity
tends to be mean-reverting, a drift term does not seem to be suitable
for intraday markets. We define the formal cost function as
\begin{equation}
C(x_{1},...,x_{N})=\sum_{k=1}^{N}n_{k}y_{0}+\sum_{k=1}^{N}\underbrace{n_{k}(\sigma_{k}\xi_{k})}_{\mathrm{volatility}}+\sum_{k=1}^{N}\underbrace{n_{k}I_{k}^{temp}(n_{k},k,t)}_{\mathrm{temporary}\,\mathrm{impact}}+\sum_{k=1}^{N}\underbrace{n_{k}I_{k}^{perm}(n_{k-1},k-1,t)}_{\mathrm{permanent}\,\mathrm{impact}}.
\end{equation}
The costs are given by the volatility of prices and the two market
impacts. In addition to just optimizing costs, \citet{almgren2001optimal}
proposed a mean-variance approach to execute volume under the constraint
of risk aversion. Based on the idea of a mean-variance optimum such
as the one in the modern portfolio theory of \citet{Markowitz}, one
can compute the expected costs $E(C)$ and its variance $V(C)$ from
Eq. (6) and derive a mean-variance optimization in 
\begin{equation}
\ensuremath{\underset{x_{k}}{\mathrm{min}}(}E(C)+\lambda V(C)),
\end{equation}
where $\lambda$ is the risk aversion parameter. The solution of Eq.
(7) yields our two novel execution approaches \textbf{Opti}\textsubscript{\textbf{C}}
and \textbf{Opti}\textsubscript{\textbf{\textgreek{sv}}}, whereas
the first model only optimizes costs and sets $\lambda=0$ while the
second assumes a certain level of risk aversion\footnote{Our choice of lambda shall ensure an appropriate level of risk aversion.
We tried more aggressive levels and found that they do not cause large
differences as more traded volume means much higher costs. The algorithm
always tries to balance the costs, even if less volatility in costs
is desired. Hence, we believe the current selection is appropriate,
as the other values do not change the path too much.} with $\lambda=2x10^{(-5)}$. Thus, we present two different models
suitable for different levels of risk appetite. However, to finally
compute a solution for Eq. (7), we need to determine the market impact
model in the next sub-section.

\subsubsection{The Choice of Market Impact Models}

Market impact models determine the mathematical complexity of Eq.
(6) to a great extent but are also the core of the execution approach
as they determine the costs. Financial markets often assume less repercussions
from individual trading and imply a linear influence, as done by \citet{almgren2001optimal}.
Exponential approaches are also utilized (see \citet{almgren2003optimal}).
Another frequently applied approach is the square-root impact model.
It is versatile and works in many diverse markets such as crypto or
options trading as \citet{toth2016square} pointed out. \citet{gatheral2012transient}
proposed a transient yet linear price impact model that decays over
time. However, there is no blueprint ready for intraday electricity
markets, which is why we follow a data-driven approach herein and
derive a numerical model based on the intraday data of the year 2018.
But how can we measure permanent and temporary market impact based
on an LOB? Permanent market impact is that which evolves after a trade
and is often caused by liquidity providers that adjust their price
levels after the arrival of transactions. As an approximation, we
first compute the mean of all trades happening within a range of 10
seconds, as shown in Figure 6.
\begin{figure}[t]
\begin{centering}
\includegraphics[scale=0.27]{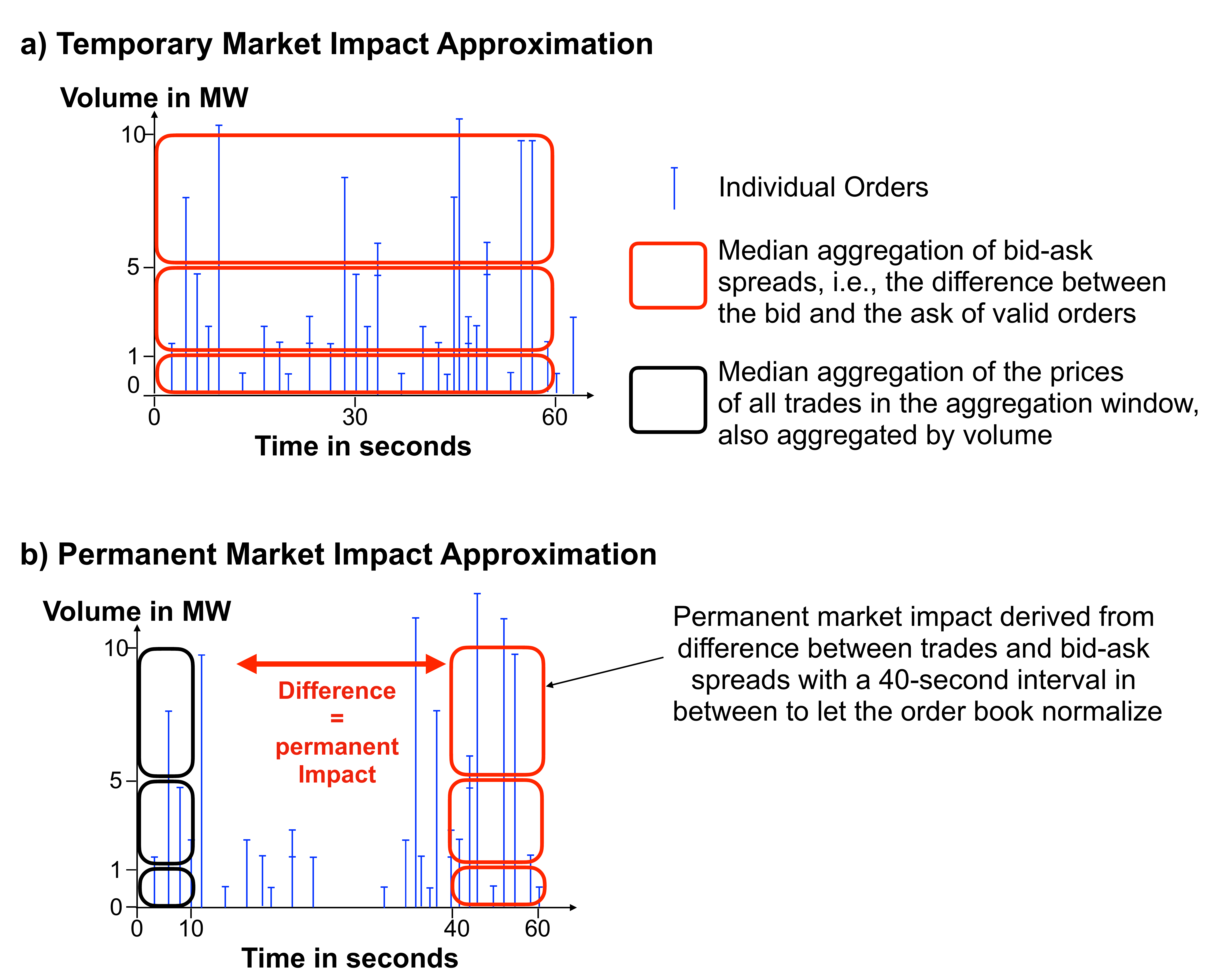}\caption{Derivation of a) temporary and b) permanent market impact from EPEX
order book data depicted in a schematic representation. The plot is
limited to the dimensions volume and time and does not show any price
levels. Obviously, all orders are sorted by prices as well, but this
is irrelevant to understanding the discretization approach. Both plots
depict a one-minute interval and highlight how the minute interval
is used to derive the impact data.}
\par\end{centering}
\end{figure}
 After another 40 seconds of time delay, assuming that the order book
normalizes again, we compute the median BAS averaged over 20 seconds
and compare its level to the aggregated BAS before the arrival of
trades. We also group the output in volume buckets to reflect that
the market impact depends on the traded volume. The resulting data
derivation is also shown in Figure 6.\\
\hspace*{0.5cm}Temporary market impact vanishes immediately after
trading and can be described as a liquidity premium that is demanded
for higher volumes in the current order book. Traders are less willing
to quote at the best possible price if the volumes are very large.
The impact is temporary, as one pays a volume premium and, afterward,
the order book volumes normalize again. Here, we derive the temporary
impact by means of the BAS per volume bucket. Based on the year 2018,
the median BAS per volume is used to derive a cost curve per minute
bucket $k$, delivery hour, and weekday (see Figure 7). As mentioned
above, we receive two cost structures that increase with higher volumes
of $n_{k}$ and less time to delivery based on the minute buckets
$k$ and\textemdash minding the seasonality of electricity prices\textemdash vary
slightly across delivery hours and weekdays (see \citet{wolff2017short}
for seasonality). We model the evolving relationship between impact
in euros/MWh and the mentioned determinants as a distribution in
\begin{align}
I_{k}^{perm}(n_{k}-1,k-1,t) & \sim\,NO(\mu,\sigma)),\\
I_{k}^{temp}(n_{k},k,t) & \sim\,NO(\mu,\sigma)),
\end{align}
where (specifically for $I_{k}^{temp}(n_{k},k,t))$
\begin{align}
log(\mu) & =\beta_{0}+\sum_{m=1}^{M}f_{m}(n_{k},k,t),\\
log(\sigma) & =\beta_{0}+\sum_{m=1}^{M}f_{m}(n_{k},k,t),
\end{align}
for $n_{k}\neq0$. If $n_{k}=0$, the impact function's result has
to be zero, as no costs occur. The above is the generalized additive
model (GAM) of \citet{hastie1990generalized} and tries to model the
parameters of a distribution\textemdash in our case the normal distribution\textemdash with
a linear model that consists of an intercept coefficient $\beta_{0}$
and $m$ additional functions $f_{m}(n_{k},k,t)$ depending on the
input factors $n_{k},k,t.$ Other applications of GAMs were supplied
by \citet{wood2017generalized} and \citet{gaillard2016additive}.
In general, a GAM tries to model distribution parameters with a linear
model that is connected closely to the well-known ordinary least squares
model. However, instead of just adding linear input variables, it
is a linear model consisting of different functions. Thus, the functions
$f_{m}$ replace the usual linear coefficients. The functions can
take a variety of forms, from simple polynomials to more complex splines.
We have tried both and found that splines model the market impact
much more precisely than polynomials or linear functions. The model
itself is fitted with the R-package \texttt{gamlss} of \citet{RigbyGAMS}
and the splines are computed using the \texttt{pb} function. We use
the log as a link function for both $\hat{\mu}$ and $\hat{\sigma}$
in our GAM. The model was determined in a limited tuning study on
2018 in-sample data using polynomials, cubic splines, and monotonic
p-splines.\\
\hspace*{0.5cm}The most simplistic form of splines are cubic ones,
meaning an ensemble of piecewise polynomials of order three that join
each other at different knots (see \citet{stasinopoulos2017flexible}
for more insight into splines in the context of GAMs). The more knots
used, the more ``fitted'' the resulting curve is. Obviously, the
choice of an appropriate spline setting is not trivial and could lead
to overfitting. A way to overcome this issue is presented by \citet{eilers1996flexible}.
Penalization removes unsuitable polynomial fits and helps to avoid
a misconstructed model. Let $\psi_{k,r}(x_{k})$ be a cubic spline
on input vector $x_{k}$ for time bucket $k$ and number of knots
$r$. Penalized splines (or p-splines) try to minimize the smoothing
parameters $\gamma$ in
\begin{equation}
\sum_{k=1}^{N}\left(y_{t}\sum_{r=1}^{R}f_{m}(x_{k,r})\right)^{2}+\sum_{r=1}^{R}\gamma\int\left\Vert f_{m}^{\prime\prime}(x_{k,r})\right\Vert _{2}^{2}dx.
\end{equation}
In a limited backtest study, p-splines modeled the impact curves of
Figure 7 best. The general goodness of fit is also reflected by the
dashed lines in Figure 5.\\
\hspace*{0.5cm}The non-linear, complex modeling of market impact
distinguishes our approach from studies like \citet{glas2020intraday},
which applied a simpler polynomial fit. The other novelty of our approach
is the explicit consideration of trading characteristics.
\begin{figure}
\centering{}\includegraphics[scale=0.25]{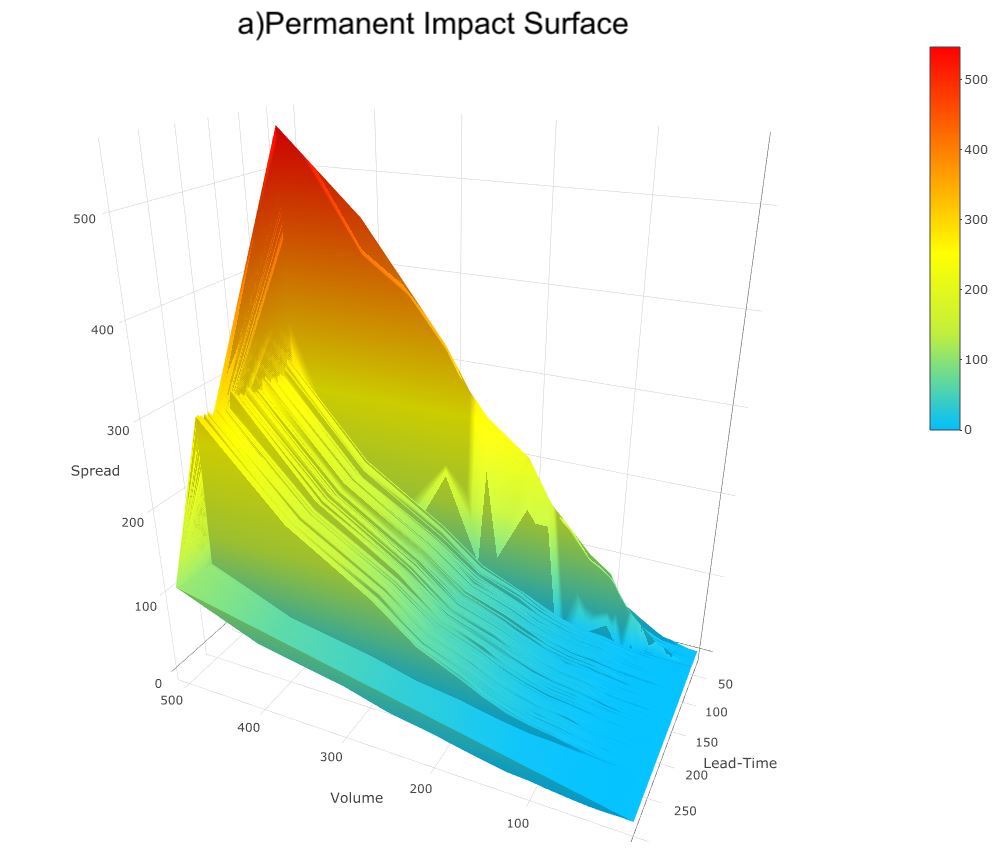}\includegraphics[scale=0.26]{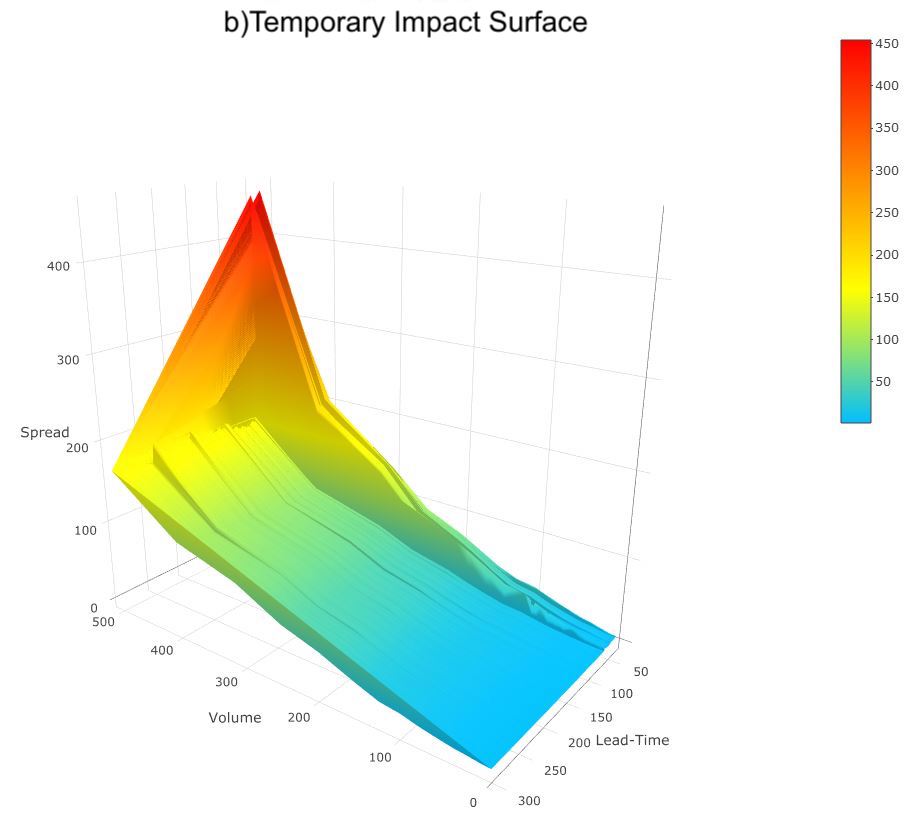}\caption{Derived cost curves for permanent and temporary market impact. The
permanent market impact is derived from BAS adjustments after trading,
and the temporary market impact derives a cost curve based on order
book depth in MW. Note that both plots assume symmetric order book
volumes as we do not divide them into buying and selling. All computations
were done with LOB data from the year 2018.}
\end{figure}
 Figure 7 shows the relationship between impact in euros per MWh and
its determinant volume (denoted as $n_{k}$) and lead-time (minute
buckets $k$) in a three-dimensional way. Figure 5 plots the relationship
in a more detailed way. Another important aspect to note is the striking
increase around 60 minutes before delivery. This has to do with the
lack of coupling afterward. Normally, XBID is only active until one
hour before delivery, meaning that during the last 60 minutes of trading,
orders are solely German ones in the German LOB. Due to system restrictions,
most market players delete their orders and re-enter them at this
specific point in time, leading to two minutes of low liquidity and
high BAS levels. Both Figures 5 and 7 confirm that. Thus, we need
to introduce a dependency of time until delivery to realistically
model the market impact. Therefore, we split the trading session into
three regimes (as shown in Figure 5 as well): XBID, cutover phase,
and non-XBID trading in 
\begin{align*}
t_{XBID} & =k\leq N-61\\
t_{cutover} & =N-60,N-59\\
t_{local} & =k>N-58,
\end{align*}
with $N$ being the number of $k$ minute buckets. Note that we define
the cutover phase as 2 minutes, starting from 59 to 60 minutes before
delivery and ending in the minute bucket starting from 60 and ending
59 minutes prior to delivery. The data suggest that both minute aggregates
feature a special impact behavior. Combining the formal definition
of Eq. (8) as well as the timing separation from above leads to our
approximation $\hat{\mu}$ of the true parameter $\mu$ in (exemplarily
for $I_{k}^{temp}(n_{k},k,t)$: 
\begin{align}
log(\mu_{k}^{temp}) & =f_{XBID}(n_{k},k,t)+f_{cutover}(n_{k},k,t)\\
 & +f_{local}(n_{k},k,t),\nonumber \\
log(\sigma_{k}^{temp}) & =f_{XBID}(n_{k},k,t)+f_{cutover}(n_{k},k,t)\\
 & +f_{local}(n_{k},k,t),\nonumber 
\end{align}
for $n_{k}\neq0$. In the case of the permanent market impact function
$I_{k}^{perm}(n_{k}-1,k-1,t)$, we needs to consider the lagged relationship
in $(n_{k}-1,k-1$). If $n_{k}=0$, we have $\mu_{k}^{temp}=0$ and
$\sigma_{k}^{temp}=0$ since no volume means no cost of trading and
no variance of such. Thus, we have different functions for the three
trading phases in our GAM model. The different functions themselves
are given by
\begin{align}
f_{XBID}(n_{k},k,t) & =\mathbbm{1}\left\{ t_{XBID}\right\} (\beta_{0,XBID}+g{}_{1,XBID}(k)\\
 & +g_{2,XBID}(n_{k})+\beta_{1,XBID}we(t)+\beta_{2,XBID}h(t)),\nonumber \\
f_{cutover}(n_{k},k,t) & =\mathbbm{1}\left\{ t_{cutover}\right\} (\beta_{0,cutover}+\beta_{1}k+g{}_{1,cutover}(n_{k})\\
 & +\beta_{2,cutover}we(t)+\beta_{3,cutover}h(t)),\nonumber \\
f_{local}(n_{k},k,t) & =\mathbbm{1}\left\{ t_{local}\right\} (\beta_{0,local}+g{}_{1,local}(k)\\
 & +g_{2,local}(n_{k})+\beta_{1,local}we(t)+\beta_{2,local}h(t)).\nonumber 
\end{align}
Note that the variance of each model is also computed by the GAM model
in the form of estimated parameter $\hat{\sigma}.$ Due to the separation
into the three regimes, we have three different sigmas per impact
function: one for XBID, one for local trading, and one for the cutover
phase.\\
\hspace*{0.5cm}The term $g(x)$ denotes a penalized spline function
of the input variable $x$ discussed earlier in the context of Eq.
(12). The function $g(x)$ does not depend on lead time in the case
of the cutover regime, as the model requires at least four distinct
values to compute a spline. During cutover, we only have two individual
$k$ values. The functions $h(t)$ and $we(t)$ of Eqs. (15)\textemdash (17)
extract peak/off-peak hours and weekend information out of delivery
time stamp $t$ and transform it into a binary dummy variable such
that the model separates between different times.

\subsubsection{Solving the Optimization Problem}

With Eqs. (12) and (14) in mind, we can compute the expected costs
and variance of Eq. (6) in more detail: 
\begin{align}
E(C) & =\underbrace{y_{0}(n_{1}+n_{2}+...+n_{N})}_{\mathrm{intraday}\,\mathrm{price}}+\sum_{k=2}^{N}\underbrace{n_{k}\mu_{k}^{perm}}_{\mathrm{permanent}\,\mathrm{impact}}+\sum_{k=1}^{N}\underbrace{n_{k}\mu_{k}^{temp}}_{\mathrm{temporary}\,\mathrm{impact}},\\
 & =\sum_{k=1}^{N}\underbrace{n_{k}y_{0}}_{\mathrm{intraday}\,\mathrm{price}}+\sum_{k=2}^{N}\underbrace{n_{k}\mu_{k}^{perm}}_{\mathrm{permanent}\,\mathrm{impact}}+\sum_{k=1}^{N}\underbrace{n_{k}\mu_{k}^{temp}}_{\mathrm{temporary}\,\mathrm{impact}}.\nonumber 
\end{align}
Note that we assume independence of the three cost components in Eq.
(18) such that the covariance of the three terms is assumed to be
zero. Since all components are deterministic and do not depend on
a random variable, their expectation is simply the function itself.
Minding the deterministic character leads to the variance given by
\begin{align}
Var(C) & =\sum_{k=1}^{N}\underbrace{n_{k}^{2}\sigma_{k}^{2}k}_{\mathrm{price\,variance}}+\sum_{k=1}^{N}\underbrace{n_{k}^{2}\sigma_{perm}^{2}(n_{k},k,t)}_{\mathrm{permanent\,impact}\,\mathrm{variance}}+\sum_{k=1}^{N}\underbrace{n_{k}^{2}\sigma_{temp}^{2}(n_{k},k,t).}_{\mathrm{temporary\,impact}\,\mathrm{variance}}
\end{align}
Putting this all together leaves the optimization problem in

\begin{align}
\underset{n_{k}}{\mathrm{min}} & ((\sum_{k=2}^{N}\underbrace{n_{k}\mu_{k}^{perm}}_{\mathrm{permanent}\,\mathrm{Impact}}+\sum_{k=1}^{N}\underbrace{n_{k}\mu_{k}^{temp}}_{\mathrm{temporary}\,\mathrm{Impact}})+\lambda(\sum_{k=1}^{N}\underbrace{n_{k}^{2}\sigma_{k}^{2}k}_{\mathrm{price\,variance}}\\
 & +\sum_{k=1}^{N}\underbrace{n_{k}^{2}\sigma_{perm}^{2}(n_{k},k,t)}_{\mathrm{permanent\,impact}\,\mathrm{variance}}+\sum_{k=1}^{N}\underbrace{n_{k}^{2}\sigma_{temp}^{2}(n_{k},k,t))}_{\mathrm{temporary\,impact}\,\mathrm{variance}}),\nonumber \\
\mathrm{subject\,to\,} & n_{k}\geq0\,(\mathrm{in\,case\,of\,\mathrm{buying}}),\nonumber \\
 & \sum_{k=1}^{N}n_{k}=X.
\end{align}
We tried a variety of different approaches such as simulated annealing
or gradient-based non-linear optimizers and found that a genetic algorithm
(see \citet{holland1992genetic} or \citet{goldberg1988genetic})
works well in our case. It is readily implemented in the R-package
\texttt{GA} of \citet{scrucca2013ga} and\textemdash inspired by
genetic processes\textemdash mutates an initial population on a random
basis. Only the parameter combinations that generate the best values
for the objective function in Eq. (20) survive, and the process is
repeated again until no more decrease of cost is reached given a certain
amount of iterations\textemdash in our case 650.

\subsection{Simple Benchmark Strategies}

\subsubsection{Instant Order Book Execution}

The last section introduced our novel execution approach tailor-made
for intraday markets. Section 3.2 will present more general execution
strategies that serve as a performance benchmark. The first approach
is not a strategy per se but the most simplistic form of trade execution.
Instant order book execution (\textbf{IOBE}) does not require any
computation or predefined trade trajectory. Instead, an energy trader
just accepts existing orders in the order book and trades the full
volume at once. In mathematical terms, this strategy is
\begin{equation}
n_{1}=X,
\end{equation}
with $n_{2,}...,n_{N}=0$. From a technical perspective, this is equal
to a click in the EPEX trading system on either the bid or the ask
side. It implies a willingness to pay the BAS for the sake of instant
execution. Figure 7 indicates the downside of such simplicity. The
higher the volume, the higher the BAS is. Prices are less favorable,
as the trader has to accept not only the first few orders on top of
the order book but a number of deeper ones as well. This usually leads
to prices being far from the surrounding small volume trades and implies
a high degree of market impact of a singular and large volume order.
Another associated risk is given by price developments. Since the
trade is entered at a specific point in time, prices can develop adversely
after the trade. This effect is less apparent with small slicing strategies
that trade over the entire time window. As such, why do traders use
IOBE? To start, its profitability is highly dependent on the traded
volume. For small traders, IOBE might be a cost-effective way to cover
open positions. More sophisticated approaches like the ones that follow
require algorithms that allocate volumes and slice them down to multiple
trades. An application programming interface (API) connection to the
exchange as well as a designated software solution is needed, while
with IOBE trades, trading can be done manually via any exchange trading
system. Thus, there is a trade-off between trading costs occurring
through IOBE versus the additional costs caused by API connections
and IT systems.

\subsubsection{Time-Weighted Average Price Execution }

Another very common strategy is an equidistant allocation of the entire
volume on each volume bucket. \citet{almgren2001optimal} referred
to this as a minimum-impact strategy, but in the world of trading,
such algorithms are commonly denoted as time-weighted average price
(\textbf{TWAP}) execution. The idea is very simple. Regardless of
empirical volume distributions, volumes are equally distributed over
time such that $n_{1}=n_{2}=...=n_{N}$ or 

\begin{equation}
x_{k}=(N-k)\frac{X}{N},
\end{equation}
\begin{equation}
n_{k}=\frac{X}{N}.
\end{equation}
The term minimum impact might be correct at first sight since we slice
the volume into as many small pieces as possible (depending on the
number of $k=1,...,N$ trading buckets) and try to tackle the factor
market influence with position size. However, the strategy does not
need to be the one with the smallest market impact, as this disregards
the distribution of the trading volume or lead time. If one trades
equal amounts and the trading volume is distributed very unequally
over time, an order can cause a larger market impact in a low-liquidity
time frame, leading to an overall larger impact than desired. However,
Figure 5 shows that the trading volume is constant for a long time
and only changes substantially in the late trading phase, which is
why we think the strategy can still be a simple enhancement to IOBE.

\subsubsection{Volume-Weighted Average Price Execution}

The thoughts of sub-section 3.2 bring us to a closely connected strategy.
While TWAP approaches do not consider deviations in trading volumes,
volume-weighted average price (\textbf{VWAP}) algorithms explicitly
do. Papers like \citet{konishi2002optimal} propose a VWAP algorithm
that tries to minimize the difference between the actual VWAP of all
trades and the trader's own VWAP. We deliberately deviate from this
as our goal is not index replication but optimal execution. For more
information on index replication strategies and pricing of such contracts,
the interested reader might refer to \citet{gueant2014vwap}, \citet{humphery2011optimal}
and \citet{bialkowski2008improving}. Recall that the VWAP itself
is given by
\begin{equation}
VWAP_{t}=\frac{1}{\sum_{k=1}^{N}v_{k}}\sum_{k=1}^{N}y_{k}v_{k}.
\end{equation}
It describes the volume-weighted average price from reception of the
position update $i$ until time of delivery $t.$ It is impossible
to achieve VWAPs if parts of the measured time lie in the past, i.e.,
you cannot replicate a VWAP of three hours of trading only by being
active solely in the last hour. Some VWAPs do not comprise the full
trading session, which is why they are sometimes called partial VWAPs
as by \citet{narajewski2018econometric} and \citet{kath2019modeling}.
In order to trade near or equal to the VWAP, we assume
\begin{equation}
x_{k}=(N-k)\left(\frac{X}{N}F_{k}\right),
\end{equation}
\begin{equation}
n_{k}=\frac{X}{N}F_{k},
\end{equation}
where $F_{j}$ is a volume reallocation factor that should ensure
a distribution based on the traded volume per minute bucket $k$ in
\begin{equation}
F_{k}=\frac{\hat{v}_{k}}{\frac{1}{N}\sum_{k=1}^{N}\hat{v}_{k}}.
\end{equation}
The actual traded volume $v_{k}$ is not known ex ante which is why
we assume $v_{k}=\hat{v}_{k}$, where $\hat{v}_{k}$ is defined as
an estimation for the true volume $v_{k}.$ In our case, $\hat{v}_{k}$
is simply the empirical volume per bucket of the year 2018. Hence,
$F_{k}$ does not change per trading hour or day and is static throughout
the entire out-of-sample test. Also, recall that index $k$ changes
with the arrival time of each position to be traded. Thus, the allocation
factor $F_{k}$ requires a new computation if the arrival time changes.
The algorithm should ideally start to trade immediately, which is
why it can be very useful to compute a set of factors $F_{k}$ for
different choices of $k$ in advance. It also follows that

\begin{equation}
\prod_{k=1}^{N}F_{k}=1,
\end{equation}
such that sufficient volume is traded. Since the strategy allocates
higher volumes to trading phases with large traded volumes, it could
be a profitable way to execute positions.

\section{Intraday Market Execution Simulation\label{sec:4}}

\subsection{Resulting Trading Trajectories}

\begin{figure}
\begin{centering}
\includegraphics[scale=0.6]{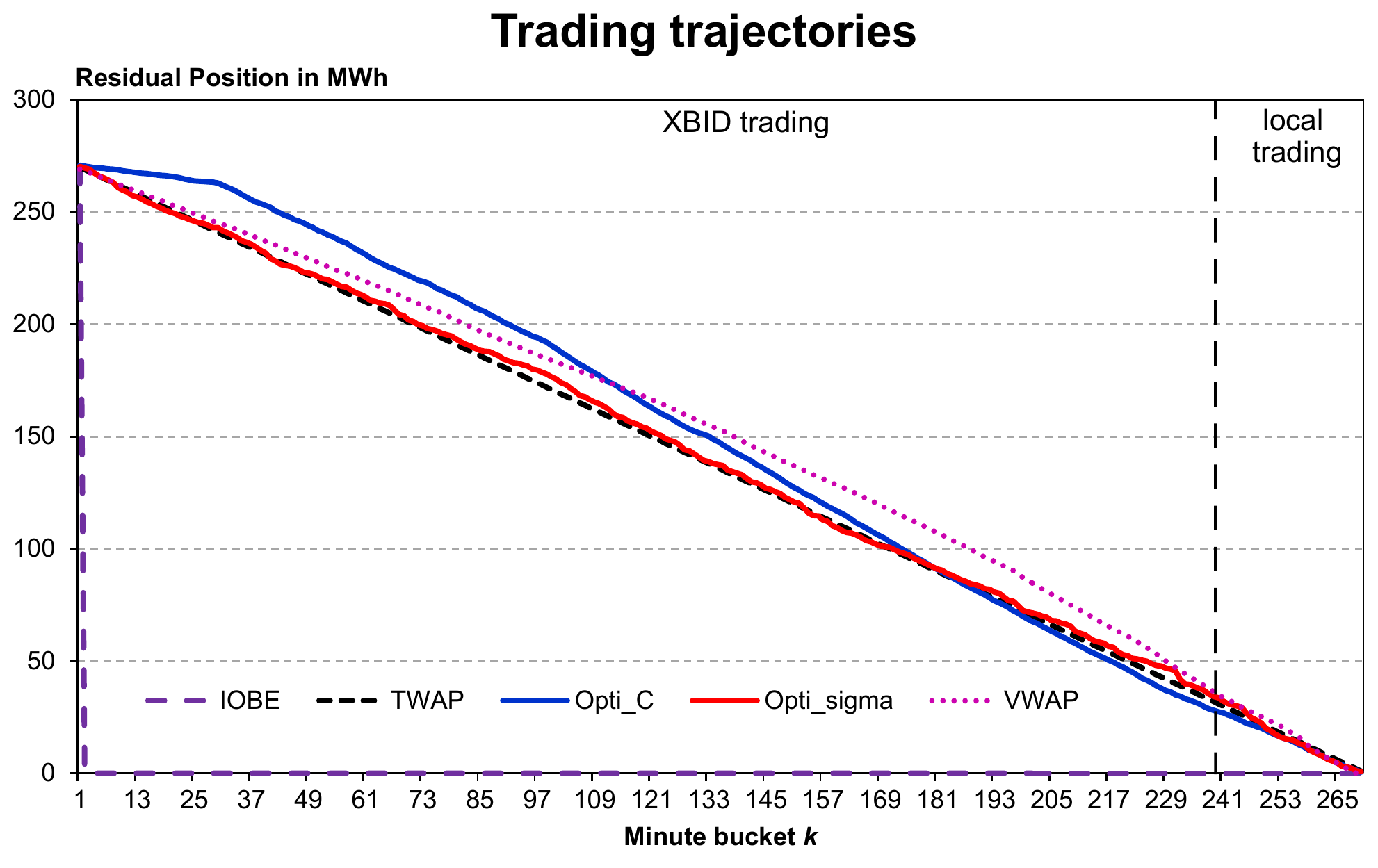}
\par\end{centering}
\caption{Trading trajectories of all algorithms of Section 3 for $k=1,...,270$,
i.e., trading from 300 to 30 minutes before delivery assuming a volume
of 270 MWh to be traded. Note that we have decided to plot the trajectory
for peak-load and weeks in case the optimization algorithms Opti\protect\textsubscript{C}
and Opti\protect\textsubscript{\textgreek{sv}}, weekend, or off-peak
trajectories partially differ. However, the overall difference is
negligible.}
\end{figure}
Different algorithms result in different trading trajectories. Section
3 focused on a mathematical formulation. In addition, we want to discuss
trading behavior to sharpen the understanding of the numerical results.
The easiest way to understand volume allocation behavior is by inspecting
a trade path plot. Figure 8 shows the trading trajectories of all
algorithms. It assumes a volume of 270 MW to be executed in 270 minute-buckets\footnote{We stop trading 30 minutes before delivery as there is only intra-grid
area trading allowed from this point onwards. However, this means
that a lead time (i.e., the nominal time to delivery) of 300 minutes
results in 270 minutes of trading.}. The two optimization algorithms Opti\textsubscript{\textgreek{sv}}
and Opti\textsubscript{C} yield different paths per total position
to be executed. However, the differences are marginal, and only very
large positions lead to a slightly smoother curve. Therefore, we have
decided to plot 270 MWh as a compromise. Positions around 200\textemdash 300
MWh apply to many market participants and depict an industry-wide
optimal execution problem.\\
\hspace*{0.5cm}Here, IOBE executes the entire volume in the first
minute bucket. This is arguably the simplest form of execution and
leads to one big trade. In contrast to that, the TWAP approach equally
splits the volume in equidistant steps, which yields a straight and
linear execution reflected by a black dashed line. It serves as a
good reference path to compare all other algorithms against. The TWAPs
try to minimize the impact by means of the smallest equidistant steps
possible. This sounds rather naive but could prove profitable if we
recap the sharp exponential growth of BAS with the higher volumes
depicted in Figure 5. Hence, it is interesting to see how other approaches
deviate from this. The VWAP algorithm allocates based on historic
volume distributions over time. This leads to an interesting pattern.
The trading volume increases with time to delivery, as shown in Figure
4. The VWAP algorithm shifts a bit of volume into later trading phases.
In the middle of the selected trading window, its inventory position
is around 10 MWh higher compared to the TWAP allocations. However,
the overall change is small and does not lead to a drastic change.\\
\hspace*{0.5cm}Another interesting pattern evolves with Opti\textsubscript{C}.
Recall that this algorithm purely optimizes costs based on a GAM trained
on 2018 data. The algorithm starts with slower trading rates than
all other approaches and leaves a higher position open in the early
and mid-trading phases. A striking behavior happens shortly before
the switch from pan-European XBID trading to local German trading.
Starting at around $k=185,$ positions are closed more aggressively
compared to TWAP and all other models, which results in a smaller
open position when local trading starts. Taking a look at Figure 5,
this makes perfect sense as this decision avoids the high BAS levels
in local trading. Figure 8 makes it almost impossible to see, but
both optimization approaches avoid the very expensive cutover phase
entirely.\\
\hspace*{0.5cm}Opti\textsubscript{\textgreek{sv}} is closely connected
to the pure cost optimization but considers risk aversion in the form
of the volatility of prices and bid-ask spreads. If one recalls, that
this approach balances between expected trading costs (which means
equally sized small volumes per bucket) and the variance of expected
costs, it becomes evident that the algorithm tries to avoid additional
variance at the cost of slightly higher market impact. This leads
to an almost TWAP-like trading path. One could ask why there is no
remarkable change in the trajectories as shown by \citet{almgren2001optimal},
wherein mean-variance approaches resulted in very different trade
trajectories. The answer is given by the market impact functions.
The mentioned paper assumed a linear impact. Thus, risk aversion does
not cause the trader to pay a large premium in the form of bid-ask
spreads for the sake of less variance in costs and execution prices.
An optimization algorithm could easily shift trading volumes into
early trading phases without causing an exponential cost increase.
This is different from intraday trading and our market impact models.
Every time trading is shifted away from a late trading point to an
early one, this causes an exponential growth in costs while reducing
volatility in an exponential way as well. Our parameter choice for
$\lambda$ seems to be very conservative. Thus, Opti\textsubscript{\textgreek{sv}}
carefully balances between expected costs and volatility. However,
we found that even more aggressive choices for $\lambda$ do not create
substantially different trading trajectories. This is interesting
as intraday markets feature price spikes and a generally high level
of volatility, as mentioned by \citet{wolff2017short}. Although there
is a high level of volatility, a mathematically optimized execution
seems to deviate only partially from the benchmark path of TWAP allocation.
\begin{table}
\centering{}\includegraphics[scale=0.7]{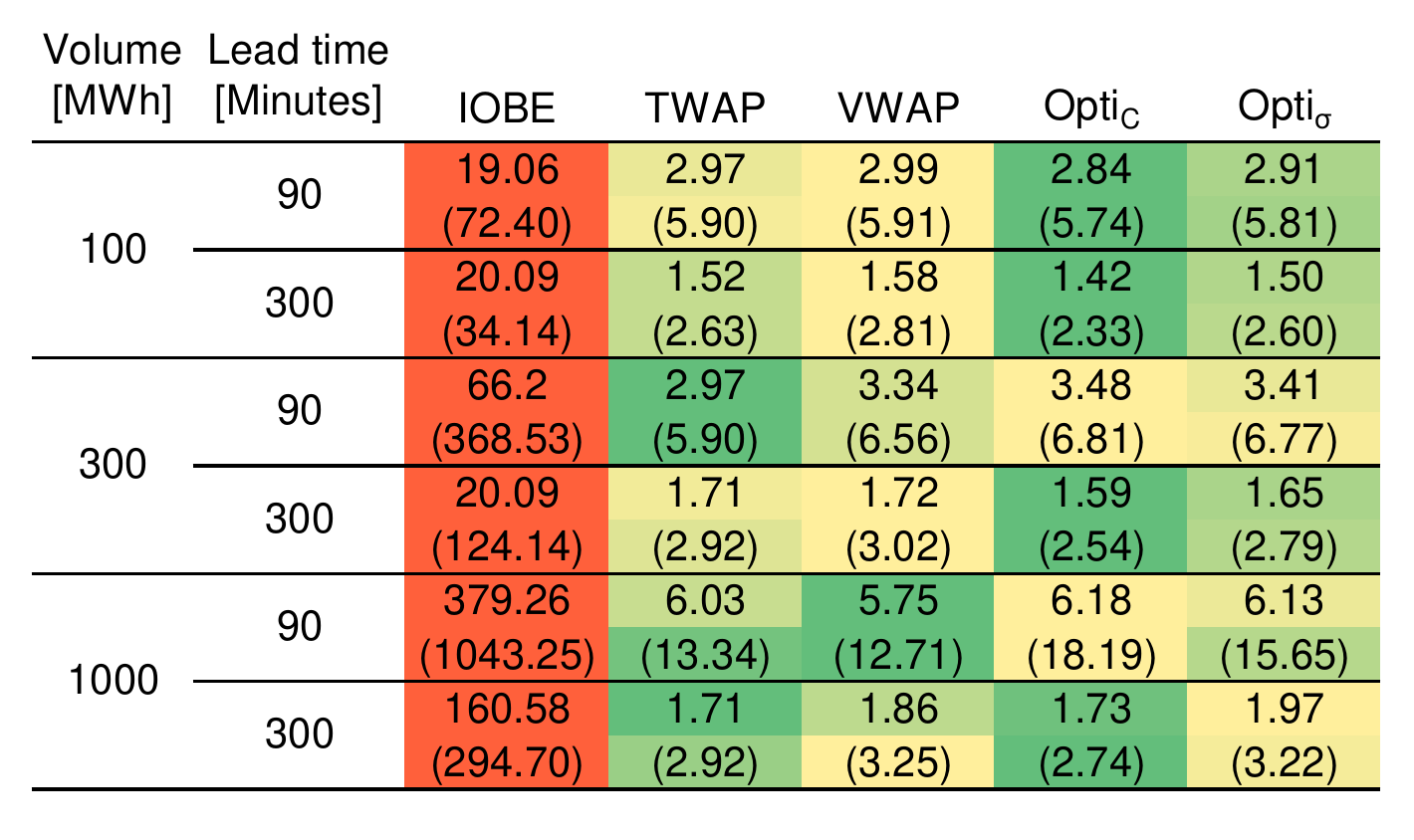}\caption{Median BAS per strategy, lead time, and volume. The lead time is connected
to the choice of $k.$ If, for instance, $k=30$, then the corresponding
lead time is 60 minutes to delivery. The unit of the Bid-Ask spread
is €/MWh. In addition to the median, we report mean values in brackets.}
\end{table}
\\
\hspace*{0.5cm}Despite the fact that we have a diverse set of algorithmic
trading models that differ in trading, computational complexity, and
consideration of market impact, the trading trajectories are less
heterogeneous than expected. In fact, this leads to the first interesting
conclusion. In trading areas with sparse order books and exponential
growth in the temporary and permanent market impact, the predominant
strategy aspect is to trade as little as possible per aggregation
step to reduce the impact and only partially deviate from that rule
of thumb for the sake of risk aversion or additional knowledge about
market regimes or volumes.

\subsection{Realized Execution Costs and Variance }

\begin{wraptable}{O}{0.5\columnwidth}%
\centering{}\includegraphics[scale=0.7]{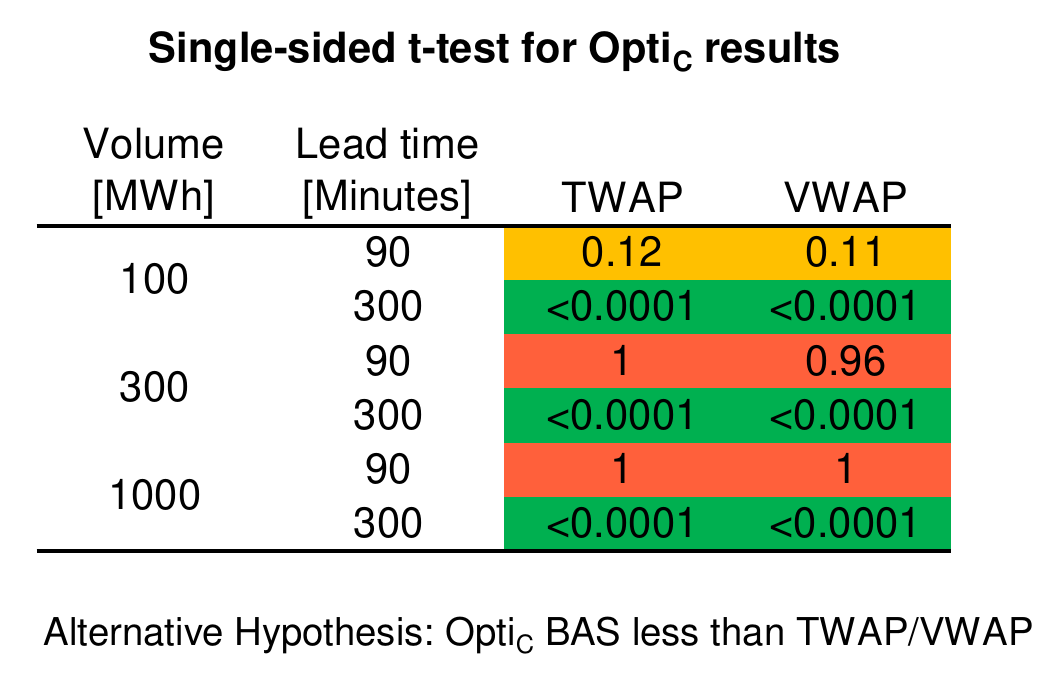}\caption{Results of a single-sided t-test testing for statistical significance
of the alternative hypothesis that Opti\protect\textsubscript{C}
results are lower than the compared alternatives.}
\end{wraptable}%
Section 4.1 demonstrated how considering different aspects such as
market coupling leads to different execution paths. It is important
to note that all models were solely trained on 2018 data. Thus, all
assumptions and optimizations might turn out to be incorrect for the
year 2019. Following the aggregation logic of section 2.2, we compare
how each algorithm trades volumes and benchmark the trades against
the aggregated BAS and prices of each individual minute bucket of
2019. The aggregation averages out all spikes and outliers while still
being very granular so that we have detailed results. Apart from that,
we analyze different trading volumes (100, 300, 1,000) and different
times to contract expiry, i.e., $k=\{60,270\},$ to reflect the behavior
under changing conditions. To our knowledge, this paper is the first
one that reports out-of-sample results for the German intraday market.\\
\hspace*{0.5cm}We want to first focus on realized execution costs
in the form of BAS, more formally defined as 
\begin{equation}
BAS_{t,k,r}=(y_{t,k,r,sell}-y_{t,k,r,buy}).
\end{equation}
Hence, the BAS is aggregated per delivery date $t$, minute bucket
$k,$ and accumulated volume $r$. Using it as a benchmark also covers
the most important trade execution aspects mentioned by \citet{perold1988implementation}.
Table 2 grants a first overview of our empirical simulation. Since
the mean is not a robust measure in the case of outliers, we have
decided to focus on the median and only report the mean in brackets
for the sake of completeness. The first striking observation is the
high level of BAS if IOBE execution is concerned. It becomes obvious
that actively ``clicking away'' the entire volume at once leads
to a dramatic cost level that no company should aim for. Simple approaches
like TWAP that do not require massive computation already beat IOBE
by far. The VWAP strategy is also much better than IOBE but stays
behind TWAP in most cases. This is somehow unexpected and shows that
a huge trading volume does not necessarily go hand in hand with low
spreads. \\
\hspace*{0.5cm}The dedicated optimization algorithm Opti\textsubscript{C}
performs as expected when the volumes per bucket are low and beats
all other approaches. However, this is only the case if there is lots
of time to close positions, i.e., when the lead time is 300 or when
there is less volume to be traded. But why is that so? We can think
of two reasons. Taking a closer look at Figure 5, we can see that
the dashed black line is below the actual spreads in the late trading
phases before XBID. That means that the costs in local trading are
slightly overestimated. Our choice of 60 minutes to trade seems to
underline the model limitations. The optimization model exhibits weaknesses
if applied in late trading and forced to trade large volumes. The
other explanation is more pragmatic. We have used 2018 data to train
the models, but that does not necessarily mean that identical patterns
will evolve for 2019. Parts of the differences might also be due to
the random noise that is present with all empirical approaches.\\
\hspace*{0.5cm}Opti\textsubscript{\textgreek{sv}} expands Opti\textsubscript{C}
by a certain level of risk aversion that flattens the trading volumes.
Consequently, there is a shift in performance. In phases where Opti\textsubscript{C}
performs well, a premium in the form of slightly lower performance
is paid. In the case of Opti\textsubscript{C} suffering from limitations,
the risk-averse approach Opti\textsubscript{\textgreek{sv}} loses
less. All models struggle to keep their spread level when volumes
increase. This observation makes sense, as the resulting orders hit
deep layers in the order book and more spread is paid. There is no
difference between the TWAP results in the cases of 100- and 300-MW
volume for a lead time of 90 minutes. While this is contradictory
at first, a deeper look gains more insight. We have aggregated into
time and volume buckets, and 300 MW/100 MW allocated into 60 time
buckets means trades of 5 MW and 1.1 MW each. It is important to note
that our second volume bucket spans from 1 to 5 MW, meaning that both
orders result in the same out-of-sample spread simply due to our aggregation
logic being too gross in that specific scenario. That is the reason
why some algorithms' results equal each other.\\
\hspace*{0.5cm}We can see that the optimization approach makes sense
in many scenarios. However, does that postulation hold true under
a stricter statistical analysis? A one-sided t-test that checks for
the significance of Opti\textsubscript{C} BAS levels being lower
than the TWAP and VWAP brings more clarity (see results in Table 3).
All scenarios with 300-MW positions are clearly significant. The results
for 100 MW and 90 minutes of lead time are a bit better but do not
persist under our stricter framework. The differences could also be
random-based. Higher volumes and 90 minutes of lead time are clearly
not significant, which further underlines the inherent Opti\textsubscript{C}
model's limitations. But what does that mean in terms of monetary
effects? Our optimization model outperforms all other models by around
15 ct/MWh if we stick to the more robust median BAS as our measure
of choice. Say a trader usually trades 100 MW per day and hour. As
an approximation for the premium on either a buy or sell trade, we
assume the half-spread, i.e., $BAS_{t,k,j}/2$. We acknowledge that
this assumption requires symmetric savings for buy and sell trades,
but for simplification matters, this should suffice. The trader will
then save 100 MW{*}24 hours{*}365 days{*}8 ct, the latter value being
roughly the difference between Opti\textsubscript{C} and TWAP. The
approach saves 70,080 euros per year due to an optimized execution.
Assume a volume of 300 MW 300 minutes before delivery, the alternatives
being IOBE and the optimized model. This comparison results in savings
of 2,775 euros for each individual delivery hour and underlines that
optimal execution can be an important performance driver.\\
\hspace*{0.5cm}Last but not least, the variance of cost requires
discussion. Equations (19) and (20) reflect how Opti\textsubscript{\textgreek{sv}}
balances between minimizing cost and variance. Inevitably, mean-variance
approaches introduce subjectivity since the risk-aversion parameter
$\lambda$ is chosen by the user. One could argue that with just a
few hours of trading, variance does not play a major role. If equity
trading is concerned, there are usually much longer holding periods
and daily volatility reports of open position, but in intraday trading,
positions are closed within minutes or hours. However, we want to
inspect if Opti\textsubscript{\textgreek{sv}} achieves the goal of
lower variance. Our measure of choice is the weighted standard deviation
of BAS, defined as 
\begin{equation}
Std^{BAS}=\sqrt{\frac{\sum_{t=1}^{T}(BAS_{t,k,r}-\mu_{t,r})^{2}}{(\frac{T-1}{T})\sum_{t=1}^{T}v_{t,k,r}^{algo}}},
\end{equation}
where $T$ equals the length of our backtest time series and $v_{t,k,r}^{algo}$
the volume allocated by each algorithm per delivery date $t$, minute
bucket $k,$ and accumulated volume $r$. It is crucial to conduct
a volume-weighting as the volumes are not distributed symmetrically.
For instance, IOBE trades just once but with 100\% of the position.
Without taking volumes into consideration, the results would be biased.
Table 4 demonstrates a mixed performance. In some cases, the variance
of Opti\textsubscript{\textgreek{sv}} is low or even the lowest across
all models. However, the difference is small and comes at the price
of a higher BAS. The overall trade trajectory is smoother and volumes
are split more equally if we compare Opti\textsubscript{\textgreek{sv}}
with Opti\textsubscript{C}. Unfortunately, both models share the
same shortcomings. Opti\textsubscript{\textgreek{sv}} does not deliver
less variance in problematic scenarios (low lead time and large positions).
All in all, the goal of lower variance is achieved. However, the trade-off
between higher spreads and lower variance does not seem to be too
convincing in intraday markets.
\begin{table}
\begin{centering}
\includegraphics[scale=0.7]{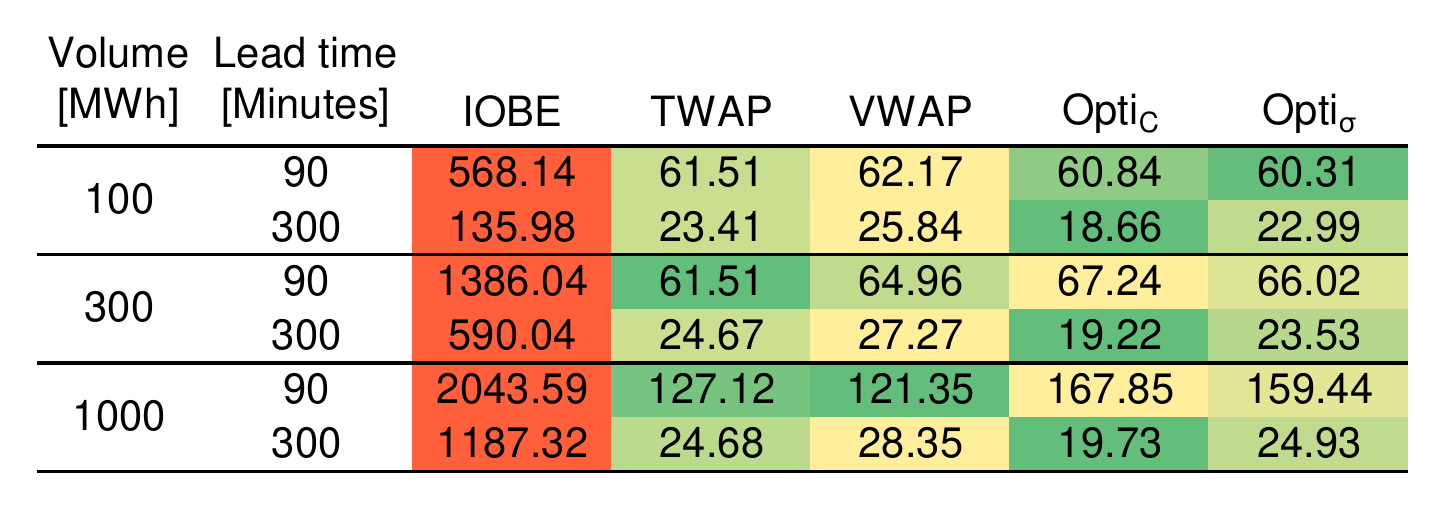}
\par\end{centering}
\centering{}\caption{Volume weighted standard deviation of BAS per strategy, lead time,
and volume. The lead time is connected to the choice of $k.$ If,
for instance, $k=30$ then the corresponding lead time is 1 hour before
delivery. The unit of the Bid-Ask spread is €/MWh.}
\end{table}

\subsection{Trade Price Benchmarking and Randomization of Volumes}

The median BAS is a suitable measure as it allows for a compressed
overview of performance. However, it features two drawbacks. First,
the BAS is a symmetrical measure and does not contain any information
on buy- or sell-side performance. If a company is a net seller, looking
at spreads has limited value. Second, spreads are not as easily interpretable
as prices, for instance. Many other papers utilize plain prices or
averages like the ID3 or ID1 as a reference. We want to exploit these
two aspects and, additionally, benchmark our algorithms against buy
and sell prices. One might argue that the BAS is a direct result of
prices, as shown in Eq. (29). This is basically true, but for this
second benchmark, we have computed the prices from scratch following
the logic of section 2.2. The findings of the price results in section
4.3 and BAS tables of section 4.2 might slightly differ due to different
roundings, i.e., a price result will not 1:1 add up to the BAS results.
While this could be perceived as confusing, we see it as an additional
layer of verification as we apply a newly computed measure.\\
\hspace*{0.5cm}Table 5
\begin{table}
\includegraphics[scale=0.6]{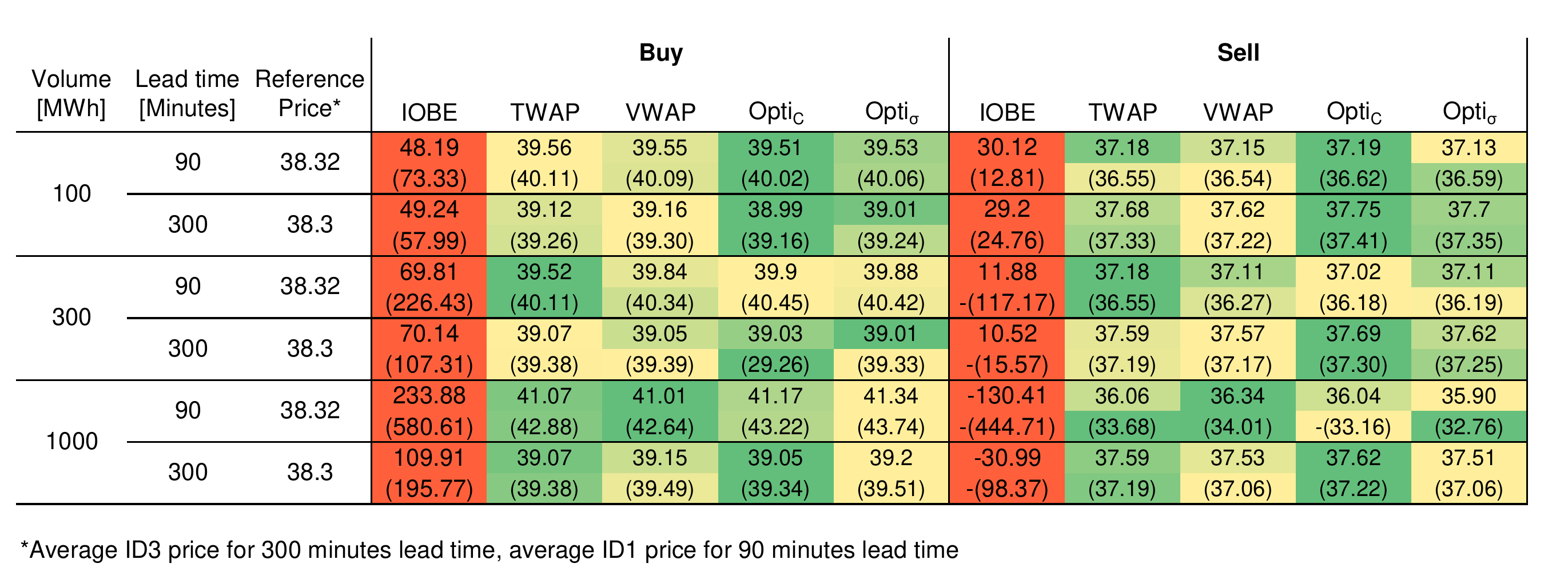}
\centering{}\caption{Average BAS per strategy, lead time, and volume. The lead time is
connected to the choice of $k.$ If, for instance, $k=60$, then the
corresponding lead time is 90 minutes before delivery. The unit of
the Bid-Ask spread is €/MWh.}
\end{table}
 reports all findings separated into buy and sell as well as the known
scenarios with different trading volumes and lead times. In addition,
we report two volume-weighted reference prices published by EPEX Spot
SE. For a lead time of 300 minutes, that is, the ID3, for 90 minutes
until delivery, we apply the ID1. These indices do not comprise the
relevant time frames 1:1 but serve as an appropriate and reproducible
benchmark. Note that we only report the indices as a reference point
but not as a measure per se. For readers interested in index-tracking
or VWAP pricing, we refer to \citet{frei2015optimal} and \citet{cartea2016closed}.
Unsurprisingly, Table 5 confirms the overall impression of our spread
analysis. Opti\textsubscript{C} yields convincing results with longer
lead times, or\textemdash if forced to liquidate positions with shorter
lead times\textemdash only in the case of small values. Moreover,
TWAP also shows favorable prices.\\
\hspace*{0.5cm}Taking a deeper look at buying and selling separately,
some interesting patterns emerge. In low-volume scenarios, the benefits
of our optimization approach are stronger when buying volumes. This
effect cancels out with higher volumes. In the case of medium volumes,
one can observe slightly better sell than buy prices (e.g., considering
the difference between Opti\textsubscript{C} and TWAP). However,
this effect is negligible. Overall, the LOB imbalance in terms of
prices is rather low. It does not seem to make a difference if one
is buying or selling, which delivers evidence to the fact that quoted
volumes and resulting execution prices are symmetrical. There might
be times where this conclusion does not hold anymore, such as large
plant outages or renewables forecast updates that cause the entire
market to sell generated power. However, these go beyond our considerations
and obviously seem to cancel out with a sample size of almost one
year of out-of-sample data.\\
\hspace*{0.5cm}Another interesting insight is connected to reference
prices. Many intraday papers like \citet{kath2018value} just apply
index prices published by EPEX Spot. However, if we compare the realized
price levels to the reference price, the difference is striking. No
algorithm can trade at the reference price. Instead, an additional
mark-up (selling at lower prices or buying at higher ones) of around
0.70\textemdash 1.20 €/MWh is required. But why is that so? We must
keep in mind that the reference price is a mid-price between buys
and sells. So, in order to trade at or near an ID3, a trader has to\textemdash at
least partially\textemdash earn the BAS. This contradicts one of our
core assumptions, namely the active trading of positions meaning accepting
orders directly in LOBs. Our algorithms explicitly pay the BAS as
they do not want to wait for execution and face the risk of non-trading
over a longer period of time. Additionally, earning the BAS, i.e.,
waiting for others to accept their own orders, requires one to be
always on top of the order book. This aspect is a trading strategy
on its own, as one needs to define price aggression, position sizes,
and inventory risks in a different way than is done in this paper.
However, the simple comparison with reference prices reveals a major
finding with large academic impact. Even complex execution algorithms
trade far  from ID1 and ID3. Thus, the assumption of trading at index
prices is an unrealistic one or needs at least a proper discussion
of index replications algorithms and optimal execution. The topic
itself is beyond the scope of this paper, but the naive comparison
of our results and the reference prices emphasizes a core limitation
of the majority of papers on intraday trading.\\
\hspace*{0.5cm}Last but not least, we want to examine if the results
persist under changing conditions. Assuming fixed positions amid all
days and hours of the year is a strong assumption that will only account
for a number of very large market players. However, what if smaller
traders utilize the execution algorithms only occasionally? In addition
to gaining more realistic results, we introduce an additional robustness
check by manipulating the sample. Do the algorithms deliver a constant
performance, or are the results of Table 5 only due to some very specific
events? Based on the described out-of-sample tests, we can erratically
remove days from the sample to find answers to these questions. Figure
9 plots the modus operandi. Based on 100\% of the out-of-sample data,
we randomly delete 60\% and then another 40\% or, first, 25\% followed
by 50\%, which results in four different randomization paths for two
different lead-time and volume scenarios. Inspecting the results,
we find that the robustness check by sampling confirms the previous
results: TWAP and VWAP are generally more profitable in high-volume,
low-lead-time scenarios, while the optimization approaches show superior
performance if given more time to trade. Thus, we have delivered evidence
of the fact that the results of Tables 2 and 5 hold true under more
general conditions. They are valid for the entire data set as well
as randomly sampled parts of it, which renders them as a good line
of orientation for market players.
\begin{figure}
\begin{centering}
\includegraphics[scale=0.55]{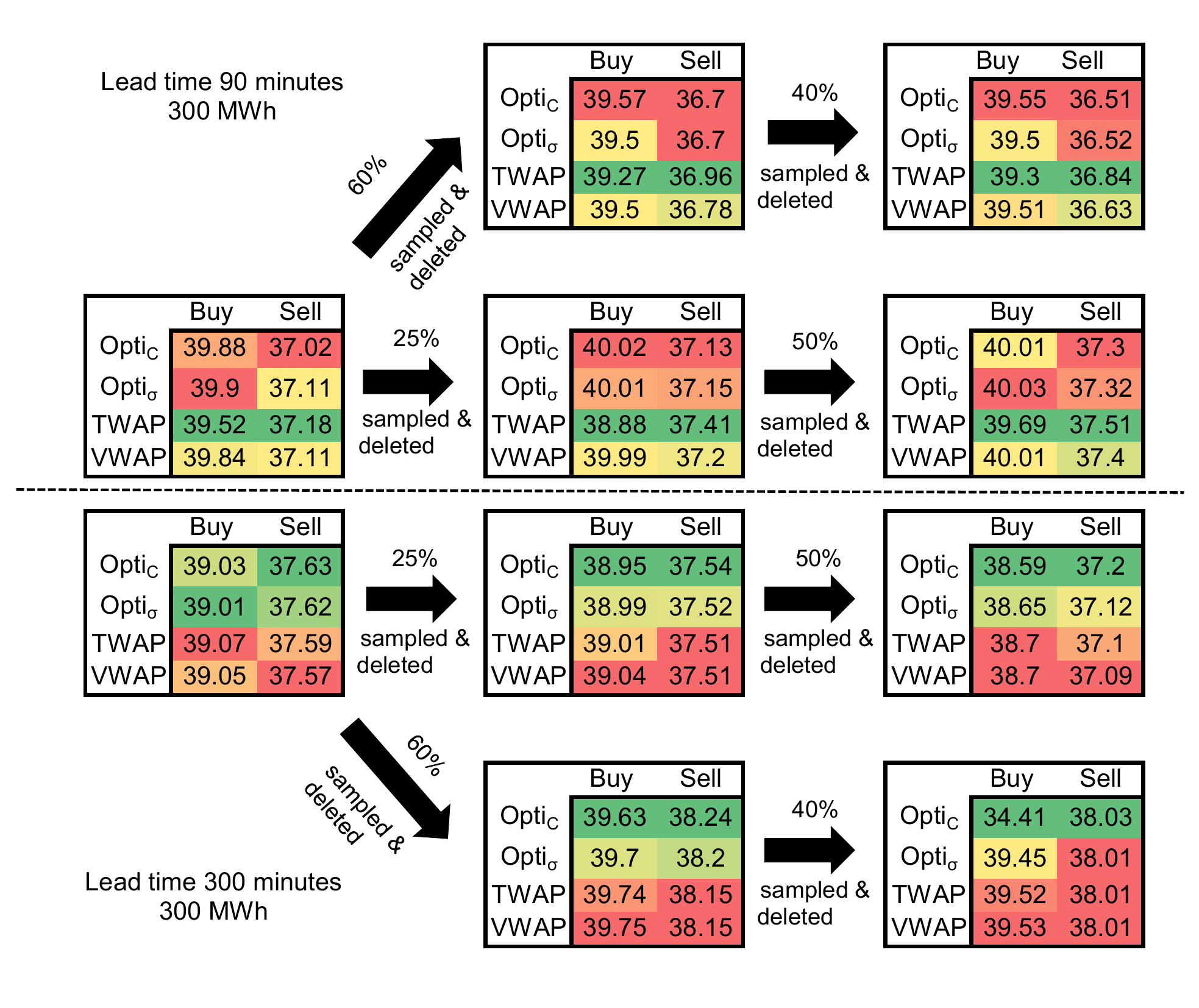}\caption{Out-of-sample scenario analysis given randomly sampled days of the
year 2019. The calculation emphasizes whether the results persist
under changing conditions or erratic occurrences of positions to be
executed, e.g., plant outages.}
\par\end{centering}
\end{figure}

\section{Contributions and Outlook\label{sec:5}}

\subsection{Conclusion}

Optimal execution of positions in intraday markets, despite its importance
with regard to a trader's performance, has only come to academic attention
recently. Recent papers like \citet{glas2020intraday} use LOB data,
aggregate information into discrete time steps, and solve numerical
optimization problems, aiming for an appropriate trade trajectory
that minimizes bid-ask spreads. Equity markets supply a rich body
of literature but lack certain characteristics like different trading
regimes, short lead-times, and exponential growth of costs due to
liquidity restrictions. We provide a more refined optimization approach.
Our model minimizes expected costs as well as\textemdash if desired\textemdash expected
variance and outputs an optimal trading path per minute.\\
\hspace*{0.5cm}Based on aggregated minute buckets, we have fitted
two GAM models to consider permanent and temporary market impact.
Our calculations have shown that penalized splines work well for modeling
the relationship between spreads and lead time, trading volumes, and
times of trading in German continuous intraday markets. Thus, the
GAMs comprise multiple p-splines and model the exponential growth
of median bid-ask spreads depending on the aforementioned variables.
Another important thing unprecedentedly considered by our approach
is European market coupling. A simple lead time analysis showed that
BAS levels dramatically increase when the European coupling of intraday
orders under XBID stops one hour before delivery. Although trading
costs normalize shortly after this cut-over phase, their overall level
tends to be higher without coupling. Our approach takes this chronological
characteristic into account and models XBID trading, local German
trading, and the cut-over phase separately. To our knowledge, it is
the first optimization method to incorporate so many market-specific
determinants based on 60-second data.\\
\hspace*{0.5cm}But is it worth the effort? We compared the model
performance in an out-of-sample study based on the year 2019 data
against other simple benchmarks like immediate order-book execution
or time- and volume-weighted average approaches. Both measures of
choice, bid-ask spreads and plain buy and sell prices, suggest that
a more complex computation yields better execution performance in
the form of lower spreads and more favorable prices. Only in late
trading phases around 90 minutes before delivery and in the case of
large volumes to be traded in that phase does a limitation of the
GAM calculation become obvious. Costs are partially underestimated,
which causes the algorithm to trade more than is optimal in specific
time buckets. However, this only occurs when being forced to liquidate
large positions. In most scenarios, an additional saving of 10\textemdash 15
cents median BAS and 15\textemdash 20 cents/MWh in the form of favorable
prices can be realized. Apart from that, the study revealed how problematic
simple clicks in the order book can be. If volumes increase, this
form of execution can cause additional costs of over 100 euros/MWh
and should be avoided. That being said, even simple algorithms like
an equally sized split of volumes on all available time buckets performs
surprisingly well and highlight the importance of discussing the optimal
execution of trades.\\
\hspace*{0.5cm}Another crucial insight is connected to optimal execution
and average prices. The majority of current intraday trading papers
(e.g., \citet{janke2019forecasting} and \citet{uniejewski2019understanding}
to name a few) utilize volume-weighted price averages as their measure
of choice. Our results suggest that this is problematic. All algorithms
trade around 1 euro/MWh away from those prices. Using them without
discussing the costs of trading in the form of market impact and paid
bid-ask spreads makes results appear less appealing. Additionally,
we believe that the discussion of optimal execution will play an important
role in the future of electricity intraday trading as markets mature
and margins shrink. It has a direct impact on each trader's performance
and is applicable in almost every situation, whether it be a plant
outage, a large position update from customer load, or a proprietary
view on the market.

\subsection{Outlook}

The previous analysis has shown that optimal execution, whilst not
being extensively discussed in the literature yet, can be an interesting
value driver. However, what role will optimal execution play in the
future, and how could further research promote its importance? We
encourage trading-oriented forecasting approaches like that of \citet{maciejowska2020pca}
or \citet{maciejowska2019day} to discuss execution as well. Benchmarks
like ID3 are a suitable first approximation but neglect the trading
character and the fact that a trader often has to pay bid-ask spreads.
Future research should steer backtests in that direction and use order
book data whenever possible instead of trades to have more realistic
results.\\
\hspace*{0.5cm}Concerning the methodology of this paper, there are
several extensions possible. We limited our model mostly to a fixed
volume that does not change over time. While this is convenient for
many cases, typical RES generation implies erratic position changes.
Thus, the algorithm and its underlying optimization problem might
be extended to consider the random character of volumes. Another interesting
direction for further research is given by the market impact functions.
They are crucial for volume allocation and determine trading behavior
to a great extent. We have derived a suitable implementation based
on data, but other approaches could prove to be even better. The empirical
study of Section 4 revealed model limitations with larger volumes
and less time to trade. The authors assumed that the market impact
functions generally work better with low volumes. The model underestimates
the exponential effects of high trading amounts per volume bucket
in late coupled trading. Therefore, it could turn out to be beneficial
to benchmark other impact models such as the transient one of \citet{gatheral2012transient}
or the well-known and often utilized square-root model (see \citet{toth2016square}).
A general comparison of different approaches would sharpen the understanding
of intraday market microstructure. Last but not least, the present
paper limited its focus to hourly trading. Quarter-hourly continuous
trading works similarly but is\textemdash due to missing European
sister markets\textemdash not as widely coupled as its hourly opponent.
An analysis and modeling of quarter hours would be novel and could
add further insight into how intraday markets work.

\appendix
\renewcommand*{\bibfont}{\footnotesize}
\begingroup \raggedright \sloppy

\bibliographystyle{plainnat}
\bibliography{Literature}

\endgroup
\end{document}